\documentclass[aps,pre,reprint, amsmath, amssymb,superscriptaddress]{revtex4-1}
\usepackage{morefloats}
\usepackage{bm}
\newcommand{\beq}{\begin{equation}}
\newcommand{\eeq}{\end{equation}}

\usepackage{comment}

\usepackage[retainorgcmds]{IEEEtrantools}
\usepackage{graphicx,tikz,placeins}
\usepackage{mathrsfs}
\usepackage{wasysym}
\usepackage{amsmath,amssymb,amsfonts,physics}
\usepackage{color}
\usepackage{float}
\usepackage[normalem]{ulem}
\usepackage{times,txfonts}
\usepackage{nicefrac}
\usepackage[colorlinks=true,linkcolor=blue,urlcolor=blue,citecolor=blue,pdfusetitle]{hyperref}
\usepackage{physics}
\usepackage{soul}
\newcommand{\mom}[1]{\langle#1\rangle}

\newcommand{\APV}[1]{#1}

\begin{document}

\title{
\APV{Exact Mapping of Nonequilibrium to Equilibrium Phase Transitions for Systems in Contact with Two Thermal Baths}
%\APV{Mapping the transition points of equilibrium and nonequilibrium systems in contact with two thermal baths} 
%Generic properties of nonequilibrium phase transitions in contact with two thermal baths
}
\author{Iago N. Mamede}
\affiliation{Instituto de F\'{\i}sica da Universidade de S\~{a}o Paulo, 05314-970 S\~{a}o Paulo,  Brazil}
\author{Carlos E. Fiore}
\affiliation{Instituto de F\'{\i}sica da Universidade de S\~{a}o Paulo, 05314-970 S\~{a}o Paulo,  Brazil}
\author{Gustavo A. L. Forão}
\affiliation{Instituto de F\'{\i}sica da Universidade de S\~{a}o Paulo, 05314-970 S\~{a}o Paulo,  Brazil}
%\author{Vit\'oria T. Henkes}
%\affiliation{Instituto de F\'{\i}sica da Universidade de S\~{a}o Paulo, 05314-970 S\~{a}o Paulo,  Brazil}
\author{Karel Proesmans}
\affiliation{ Niels Bohr International Academy, Niels Bohr Institute,
University of Copenhagen, Blegdamsvej 17, 2100 Copenhagen, Denmark}
\author{André P. Vieira}
\affiliation{Instituto de F\'{\i}sica da Universidade de S\~{a}o Paulo, 05314-970 S\~{a}o Paulo,  Brazil}

\date{\today}

\begin{abstract}
\APV{We show that a large class of nonequilibrium many-body systems in contact with two thermal baths admit an exact mapping onto equivalent equilibrium systems. This mapping provides direct access to nonequilibrium phase transition points from known equilibrium results, irrespective of the model, interaction topology, or distance from equilibrium. We verify the universality of this correspondence using paradigmatic models (Ising, Potts, and Blume-Capel), and highlight distinctive features in entropy production close to critical and tricritical points. Our findings connect equilibrium and nonequilibrium statistical mechanics, with implications for microscopic thermal machines and stochastic thermodynamics.}
\end{abstract}

\maketitle

{\it Introduction---}
\APV{Understanding nonequilibrium phase transitions remains one of the central challenges in modern statistical physics. While their equilibrium counterparts are fully described by the optimization of a thermodynamic potential, a universal framework for nonequilibrium phase transitions is still lacking \cite{marro2005nonequilibrium, Odor2004, Henkel2008, Blythe2003, Zakine2023}. Here we address this problem for a broad class of systems in alternating contact with two thermal baths.}

\APV{Systems coupled to two or more distinct reservoirs are a paradigmatic setting to explore nonequilibrium physics, serving as canonical models for phenomena ranging from biology-inspired problems, to heat conduction in nanoscale junctions, to the operation of quantum engines \cite{danielPhysRevResearch.2.043257, Kosloff2014, Segal2016, Niedenzu2018, Cangemi2024}. While previous studies of phase transitions in such systems have often relied on approximations or focused on specific solvable cases, they have hinted at connections between equilibrium and nonequilibrium behaviors \cite{Tome_1991, Tome2012, Aron2020, Herpich2020, martynec2020entropy, aguilera2023nonequilibrium, Yan2023, Tome2023}. Our work moves beyond these specific examples by providing an exact and general mapping, demonstrating that for a vast class of interacting systems, the critical manifold is not just analogous to, but formally identical to that of an equilibrium counterpart.}

\APV{
In this Letter, we show that for systems coupled to two thermal baths, the phase diagram and critical exponents are identical to those of a corresponding equilibrium system governed by effective interaction parameters, which we derive explicitly. We show that the nonequilibrium steady-state (NESS) distribution can be written in exact closed form, independent of the microscopic model, lattice topology, or distance from equilibrium. Most importantly, the location of nonequilibrium phase transitions is obtained by a direct mapping from the corresponding equilibrium singularities. These results are valid in the limit of fast alternating contact with the baths, but we give evidence that they still approximately hold when the alternation rate is finite. 
}

\APV{
We demonstrate the generality of these results in nonequilibrium extensions of three paradigmatic models—the Ising, Potts, and Blume-Capel systems—covering continuous, discontinuous, and tricritical transitions. Alongside standard order parameters, we analyze entropy production, showing its nonuniversal character \cite{martynec2020entropy}  and revealing characteristic signatures near nonequilibrium tricriticality. This framework provides a universal bridge between equilibrium and nonequilibrium statistical mechanics, and is directly relevant for the design of microscopic thermal machines.
}

\begin{figure*}
    \centering
    \includegraphics[width=1\textwidth]{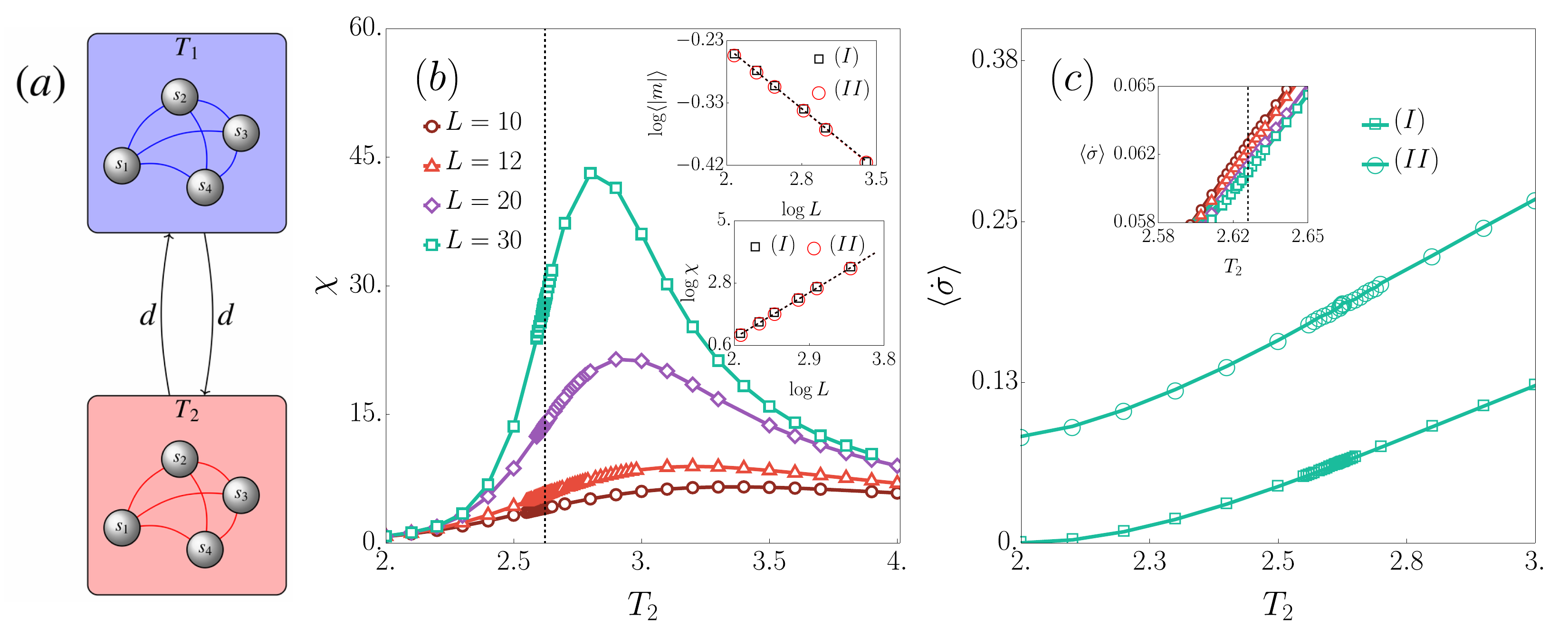}
     \caption{
     \APV{(a) An illustration of a setup for the class of systems considered here, containing $N=4$ units with interaction parameters (represented by colored edges) whose values may depend on the contact with either of two baths at temperatures $T_1$ and $T_2$.  %\CEF{(a).  It approaches to the non simultaneous description in which the system is considered at two different temperatures but there is a  large rate $d$ of switchings between thermal reservoirs, as $d$ increases (b).}. 
     Center and right: simulation results for the $L\times L$ square-lattice Ising model, with various system sizes $L$, under both field combinations (see main text). (b) Main plot shows the order-parameter variance $\chi$ versus $T_2$, while insets show, in log scale, the size-dependence of the order parameter $\langle |m|\rangle$ and its variance $\chi$ at criticality.  Dashed lines in the insets have slopes consistent with Ising critical exponents. (c) Main plot shows the entropy production $\langle {\dot \sigma}\rangle$ per site versus $T_2$ for $L=30$, while inset shows no evidence of a singular behavior for $\langle {\dot \sigma}\rangle$ at criticality. Parameters: $\mathcal{J}^{(1)}=\mathcal{J}^{(2)}=1$, $T_1=2$; for field combination (II), $h^{(1)}=-1/2$. 
     %DEIXAR APENAS A ATUAL FIGURA (b), GIRADA PARA QUE OS RESERVATÓRIOS FIQUEM UM MAIS ACIMA E OUTRO MAIS ABAIXO. MUDAR AS INDICAÇÕES DAS FIGURAS PARA (a), (b) E (c). MUDAR O RÓTULO DO INSET NA ÚLTIMA FIGURA PARA \(T_2\). CONSIDERAR REPETIR A LEGENDA DOS TAMANHOS NA ÚLTIMA FIGURA.
     }
     }
     \label{fig1}
\end{figure*}

{\it Model description and general phase-transition behavior--}
\APV{
As sketched in Fig.~\ref{fig1}, consider a system composed of \(N\) units, whose individual states can be represented by discrete values, so that the microscopic configuration of the system corresponds to \(s\equiv(s_1,s_2,\ldots,s_N)\), in which \(s_j\) represents the state of unit \(j\). For instance, in a spin-1/2 Ising model \(s_j\in\{-1,+1\}\), whereas for the $q$-state Potts model $s_j\in\{0,...,q-1\}$. We assume that the system alternates with rate \(d\) between two thermal baths   with different inverse temperatures \(\beta_1\equiv1/k_BT_1\) (``cold'') and \(\beta_2\equiv1/k_BT_2\) (``hot''). For the sake of generality, when in contact with bath \(\nu\) (\(\nu\in\{1,2\}\)), the individual elements in the system interact via an energy function \(E^{(\nu)}(s)\), characterized by a parameter vector with \(P\) components, 
\begin{equation}
\boldsymbol{\epsilon}^{(\nu)}\equiv\left( \epsilon^{(\nu)}_1,\epsilon^{(\nu)}_2,\ldots,\epsilon^{(\nu)}_P \right),
\end{equation}
% Essa descrição abaixo está incorreta. Os parâmetros rotulados por i não fazem referência à unidade i.
%where each $\epsilon^{(\nu)}_i$ encompasses the set of parameters (individual and interaction energies) for the unit $i$.
We write the energy of the system so that the parameters couple linearly with functions of the various \(s_j\).
For instance, in a spin-1/2 Ising model whose energy is given by \(E^{(\nu)}(s) = - \mathcal{J}^{(\nu)} \sum_{\langle i,j\rangle}s_is_j-h^{(\nu)}\sum_i s_i\), the two parameters represent the exchange constant between neighbor spins, \(\mathcal{J}^{(\nu)}\), and the magnetic field, \(h^{(\nu)}\), so that \(\boldsymbol{\epsilon}^{(\nu)}=\left( \mathcal{J}^{(\nu)}, h^{(\nu)} \right)\). In an experimental setup, it would be easier to control the values of parameters (such as a magnetic field or a chemical potential) coupling linearly to single units, but here we will work in the most general case.
}

%
%Although a natural choice of parameters is carried out by  varying temperatures and individual energies (e.g. in analogy with variation of the magnetic field or chemical potential) for interaction energies held fixed, we shall consider the most general case in which it can assume different values. 
\APV{
We assume that, when the system is in contact with a given bath, the dynamics involves changes in the local state of a single unit at a time, say unit \(j\), thus connecting configurations \(s\equiv (s_1,\) \(\ldots,s_{j-1},s_j,s_{j+1},\) \(\ldots,s_N)\) and \(s^\prime\equiv (s_1,\) \(\ldots,s_{j-1},s^\prime_j,s_{j+1},\) \(\ldots,s_N)\). Following a ``two-box'' description \cite{danielPhysRevResearch.2.043257, liang2021dissipation, busiello2021dissipation, mamede2025collective}, according to which the system is in contact with a single bath at a time, we assume a constant stochastic rate \(d\) of contact alternation. In the fast alternation limit, in which \(d\) is much larger than the rates of local changes, if \(p_s(t)\) is the probability that the system is in configuration \(s\) at time \(t\), the time evolution of \(p_s(t)\) is governed by the master equation \cite{danielPhysRevResearch.2.043257}
\begin{equation}
    \dot{p}_s(t)=\sum_{\nu}\sum_{s^\prime\neq s} J^{(\nu)}_{ss^\prime}(t), 
    \label{eq:pJJ}
\end{equation}
in which \(\nu\in\{1,2\}\) labels the baths and the probability currents \(J^{(\nu)}_{s s^\prime}(t)\) are given by 
\begin{equation}
J^{(\nu)}_{ ss^\prime}(t) = \omega^{(\nu)}_{s s^\prime} p_{s^\prime}(t) - \omega^{(\nu)}_{s^\prime s} p_s(t).  
\end{equation}
We also assume that the transition rate \(\omega^{(\nu)}_{s s^\prime}\) between configurations \(s^\prime\) and \(s\) takes the common Kramers form
\begin{equation}
\omega^{(\nu)}_{s s^\prime} = \frac{1}{2}\Gamma e^{-\frac{1}{2}\beta_\nu Q^{(\nu)}_{s s^\prime }},
\label{eq:Kramers}
\end{equation}
in which \(Q^{(\nu)}_{s s^\prime}= E^{(\nu)}(s) - E^{(\nu)}(s^\prime)\equiv\Delta E^{(\nu)}_{s s^\prime}\) is the heat involved in the change, \(\Gamma\) fixes the time scale
and the overall factor of $1/2$ accounts for the equal switching probability between thermal baths. 
%Transition rates between configurations differing by the states of more than one unit are assumed to be zero.
From now on, we take \(\Gamma=2\) and \(k_B=1\).
}
%Defining \(p_s^{(\nu)}(t)\) to be the probability that the system is in configuration \(s\) at time \(t\) while in contact with bath \(\nu\), we write the pair of master equations
%\begin{eqnarray}
%
%\end{eqnarray}
%}
%

\APV{
Rearranging terms, Eq.~\eqref{eq:pJJ} can be written as
\begin{equation}
    \dot{p}_s(t)=\sum_{s^\prime\neq s}\left[ \left( \omega^{(1)}_{s s^\prime} + \omega^{(2)}_{s s^\prime} \right) p_{s^\prime}(t) - \left( \omega^{(1)}_{s^\prime s} + \omega^{(2)}_{s^\prime s} \right) p_s(t) \right].
    \label{eq:pAB}
\end{equation}
We seek a stationary solution \(p^{{\rm st}}_{s}=\lim_{t\rightarrow\infty} p_s(t)\) obtained when the term in square brackets on the right-hand side of Eq.~\eqref{eq:pAB} equals zero % as \(t\rightarrow\infty\) 
for all \(s \neq s^\prime\). 
% VAMOS REAVALIAR A RELEVÂNCIA DO COMENTÁRIO ABAIXO
%Such condition implies $J^{(1)}_{s^\prime s}=-J^{(2)}_{s^\prime s}$ and it is also referred as stalling, being important in different cases, such as nanoscopic electronic devices and have been used for probing physical properties of small biological systems like molecular motors \cite{PhysRevLett.117.180601}. 
This yields an extension of the detailed-balance condition,
\begin{equation}
    \frac{p^{{\rm st}}_{s^\prime}}{p^{{\rm st}}_{s}} = \frac{\omega^{(1)}_{s^\prime s} + \omega^{(2)}_{s^\prime s}}{\omega^{(1)}_{s s^\prime} + \omega^{(2)}_{s s^\prime}}=e^{-\frac{1}{2}\beta_1 \Delta E^{(1)}_{s^\prime s}-\frac{1}{2}\beta_2 \Delta E^{(2)}_{s^\prime s}},
    \label{eq:gdb}
\end{equation}
where the relation \(\Delta E^{(\nu)}_{s^\prime s} = -\Delta E^{(\nu)}_{s s^\prime}\)
was used. The NESS probability distribution then acquires the generic form
%\begin{equation}
%     \frac{p_{s^\prime}(t\rightarrow\infty)}{p_s(t\rightarrow\infty)} = \frac{e^{-\frac{1}{2}\beta_1 \Delta E^{(1)}_{s^\prime s}}+e^{-\frac{1}{2}\beta_2 \Delta E^{(2)}_{s^\prime s}} }{e^{\frac{1}{2}\beta_1 \Delta E^{(1)}_{s^\prime s}}+e^{\frac{1}{2}\beta_2 \Delta E^{(2)}_{s^\prime s}}},
%     \label{eq:psprimeps}
%\end{equation}
%\begin{equation}
%  \frac{p^{{\rm st}}_{s^\prime}}{p^{{\rm st}}_{s}}  = e^{-\frac{1}{2}\beta_1 %\Delta E^{(1)}_{s^\prime s}-\frac{1}{2}\beta_2 \Delta E^{(2)}_{s^\prime s}}.
%\end{equation}
%\begin{equation}
%    p^{{\rm st}}_{s^\prime} = \frac{1}{Z_{12}}e^{-\tilde{\mathcal{H}}_s\left( \frac{1}{2}\beta_1 \boldsymbol{\epsilon^{(1)}} + \frac{1}{2}\beta_1 \boldsymbol{\epsilon^{(2)}} \right)}
%    \label{eq:psinfH}
%\end{equation}
\begin{equation}
    p^{{\rm st}}_{s} = \frac{1}{Z_{12}}e^{-\frac{1}{2} \left(\beta_1E^{(1)}(s)+\beta_2E^{(2)}(s)\right)},
    \label{eq:psinf}
\end{equation}
where 
\begin{equation}
    Z_{12} = \sum_s e^{-\frac{1}{2} \left(\beta_1E^{(1)}(s)+\beta_2E^{(2)}(s)\right)}
    \label{eq:Z12}
\end{equation}
is a normalization factor analogous to the equilibrium canonical partition function.
%\begin{equation}
%    Z_{12} = \sum_s e^{-\tilde{\mathcal{H}}_s\left( \frac{1}{2}\beta_1 \boldsymbol{\epsilon^{(1)}} + \frac{1}{2}\beta_1 \boldsymbol{\epsilon^{(2)}} \right)},
%    \label{eq:psinfH}
%\end{equation}
%\begin{equation}
%    Z_{12} = \sum_s e^{-\frac{1}{2} \left(\beta_1E^{(1)}(s)+\beta_2E^{(2)}(s)\right)}.
%    \label{eq:psinfH}
%\end{equation}
%in which \(\tilde{\mathcal{H}}_s\left(\boldsymbol{K}\right)\), with a dimensionless parameter vector \(\boldsymbol{K}=(K_1,K_2,\ldots,K_P)\), can be interpreted as the dimensionless version of the Hamiltonian for the model describing the system in contact with a single bath at inverse temperature \(\beta\) and parameter vector \(\boldsymbol{\epsilon}\), such that \(\boldsymbol{K}=\beta\boldsymbol{\epsilon}\). 
%\begin{equation}
%    p_s(t\rightarrow\infty) = \frac{1}{Z_{12}}e^{-\frac{1}{2}\beta_1 E^{(1)}(s)-\frac{1}{2}\beta_2 E^{(2)}(s)},
%    \label{eq:psinf}
%\end{equation}
%where $Z_{12} = \sum_s e^{-\frac{1}{2}\beta_1 E^{(1)}(s)-\frac{1}{2}\beta_2 E^{(2)}(s)}$
%is a normalization factor analogous to the equilibrium canonical partition function. Indeed, it becomes a partition function in the limit \(\beta_1=\beta_2\) and \(E^{(1)}(s)=E^{(2)}(s)\). 
%Note that the stationary distribution reduces to  Gibbs distribution at thermal equilibrium. }
Indeed, it becomes a partition function in the limit \(\beta_1=\beta_2\) and \(\boldsymbol{\epsilon}^{(1)}=\boldsymbol{\epsilon}^{(2)}\). In that case, the NESS distribution corresponds to a Gibbs distribution and to thermal equilibrium. In general, however, the dynamics leads to a nonequilibrium stationary distribution, characterized by a nonzero entropy production, as it will be shown later.
}

% (André) O parágrafo do meu rascunho que foi removido definia \boldsymbol{K}, mas sem essa definição o parágrafo seguinte não fica claro. Por isso, recoloquei o parágrafo abaixo, mas mudei a notação de K para \tilde\epsilon.

\APV{
Now notice that Eqs.~\eqref{eq:psinf} and \eqref{eq:Z12} can be rewritten as
\begin{equation}
    p_s^{\mathrm{st}} = \frac{e^{-\tilde{\mathcal{H}}_s\left( \frac{1}{2}\beta_1 \boldsymbol{\epsilon^{(1)}} + \frac{1}{2}\beta_1 \boldsymbol{\epsilon^{(2)}} \right)}}{Z_{12}}, 
    \quad
    Z_{12} = \sum_s e^{-\tilde{\mathcal{H}}_s\left( \frac{1}{2}\beta_1 \boldsymbol{\epsilon^{(1)}} + \frac{1}{2}\beta_1 \boldsymbol{\epsilon^{(2)}} \right)},
    \label{eq:psinfH}
\end{equation}
%and
%\begin{equation}
%    Z_{12} = \sum_s e^{-\tilde{\mathcal{H}}_s\left( \frac{1}{2}\beta_1 \boldsymbol{\epsilon^{(1)}} + \frac{1}{2}\beta_1 \boldsymbol{\epsilon^{(2)}} \right)},
%    \label{eq:Z12H}
%\end{equation}
in which \(\tilde{\mathcal{H}}_s\left(\boldsymbol{\tilde\epsilon}\right)\), with a dimensionless parameter vector \(\boldsymbol{\tilde\epsilon}=(\tilde\epsilon_1,\tilde\epsilon_2,\ldots,\tilde\epsilon_P)\) and
\begin{equation}
\tilde\epsilon_i \equiv \frac{1}{2}\beta_1 \epsilon^{(1)}_i + \frac{1}{2}\beta_2 \epsilon^{(2)}_i\quad\left(i\in\{1,2,\ldots,P\}\right),
\label{beh}
\end{equation}
can be interpreted as the dimensionless version of the Hamiltonian for the model describing a fictitious system in contact with a single bath at inverse temperature \(\beta\) and parameter vector \(\boldsymbol{\epsilon}\), such that \(\boldsymbol{\tilde\epsilon}=\beta\boldsymbol{\epsilon}\). In this interpretation, the NESS probability distribution would be formally equal to the equilibrium distribution of the fictitious system with suitably chosen parameters.
}
% ACHO QUE DEVERÍAMOS DEIXAR A MENÇÃO ABAIXO PARA QUANDO DISCUTIMOS MODELOS DE CAMPO MÉDIO, PORQUE FOI SÓ NESSE LIMITE QUE O CÁLCULO FOI FEITO, CERTO?
%\CEF{In SM,  the equivalence between Eq.~(\ref{eq:psinfH}) and steady solution of Eq.~(\ref{eq:pAB}) calculated via the spanning tree method \cite{schnakenberg} is exemplified.}
% Tive que reescrever o resultado principal. Embora ele esteja correto quando há uma superfície crítica, ele não é válido quando há um único ponto crítico, caso do modelo de Ising de spin 1/2. Com isso, temos que deixar o resultado em uma forma vetorial, cuja expressão explícita tem que ser discutida caso a caso.
\APV{
Phase transitions in equilibrium systems are associated with singularities in the partition function. Therefore, from Eq.~\eqref{eq:psinfH}, we conclude that, if there is a choice of \(\boldsymbol{\tilde\epsilon}\) corresponding to some relation \(f(\boldsymbol{\tilde\epsilon})=\boldsymbol{0}\) that makes \(Z_{12}\) singular, then a nonequilibrium phase transition must occur for 
%Phase transitions in equilibrium systems are associated with singularities in the partition function. From Eq.~\eqref{eq:Z12H}, we conclude that, if there is a choice of \(\boldsymbol{K}=\beta\boldsymbol{\epsilon}\) corresponding to some relation \(f(\boldsymbol{K})=\boldsymbol{0}\) which makes \(Z_{12}\) singular, then a nonequilibrium phase transition must occur for \(f\left( \frac{1}{2}\beta_1 \boldsymbol{\epsilon^{(1)}} + \frac{1}{2}\beta_2 \boldsymbol{\epsilon^{(2)}} \right) = \boldsymbol{0}\).
\begin{equation}
f\left( \frac{1}{2}\beta_1 \boldsymbol{\epsilon^{(1)}} + \frac{1}{2}\beta_2 \boldsymbol{\epsilon^{(2)}} \right) = \boldsymbol{0}.
    \label{eqe}
\end{equation}
The precise form of the relation \(f(\boldsymbol{\tilde\epsilon})=\boldsymbol{0}\) depends on the model, and some examples will be given shortly.
}

\begin{figure}
    \centering
    \includegraphics[width=0.5\textwidth]{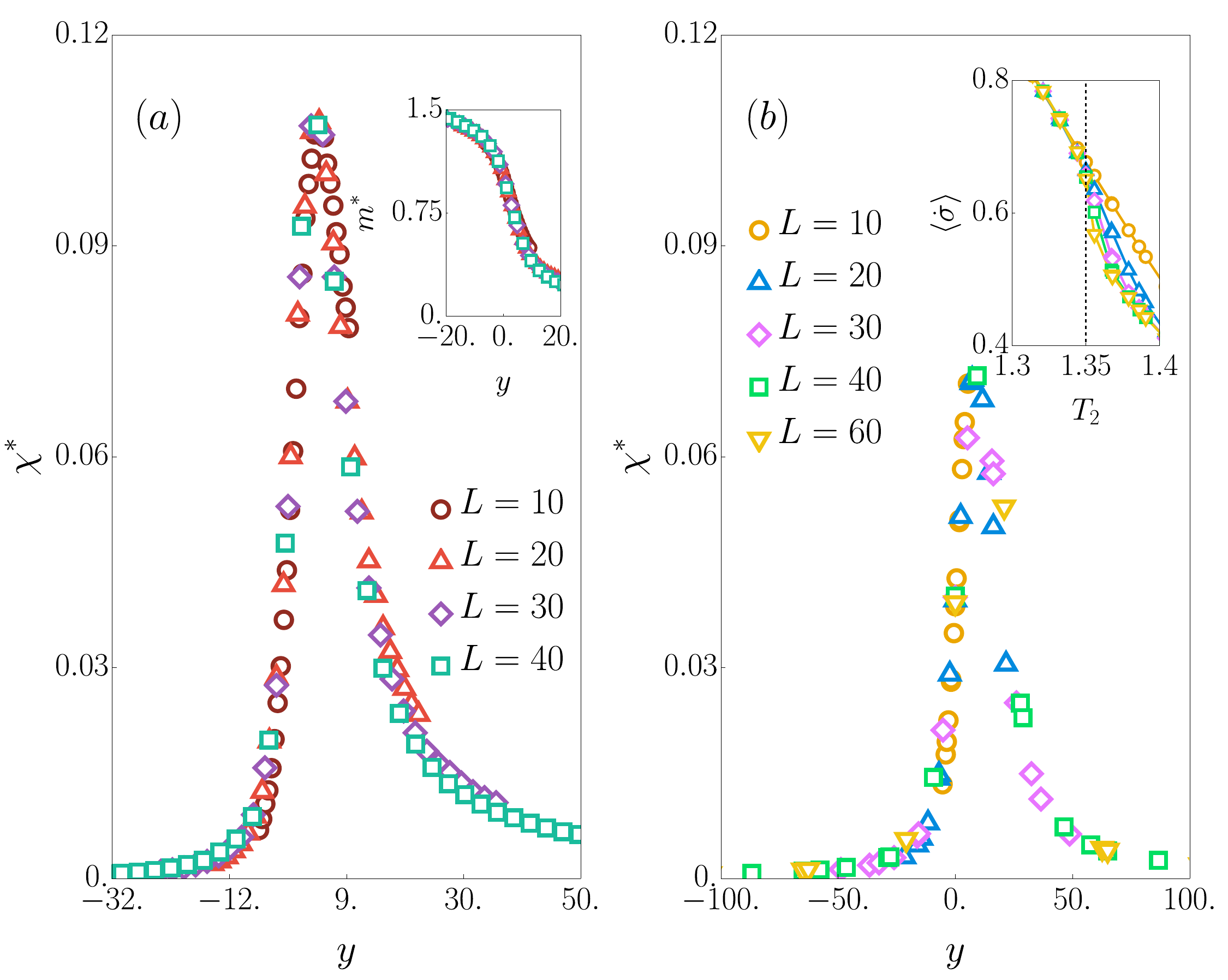}
     \caption{
     \APV{Simulation results for the nonequilibrium square-lattice \(q\)-state Potts model. The left (right) panel shows results for \(q=3\) (\(q=6\)), for which the order-disorder transition is continuous (discontinuous). Left panel: rescaled variance \(\chi^*=\chi L^{-\gamma/\nu_\perp}\) (main panel) and rescaled order parameter \(m^*=\mom{m}L^{\beta/\nu_\perp}\) (inset) versus rescaled distance to criticality $y=(T_{2}-T_{2c})L^{1/\nu_\perp}$, with $\beta=1/9$, $\gamma=13/9$, and $\nu_{\perp}=5/6$. Right panel: $\chi^*=\chi L^{-2}$ (main panel) and \(\mom{\dot{\sigma}}\) (inset) versus $y=(T_{2}-T_{2c})L^{-2}$ and  $T_2$, respectively. The critical bath temperature \(T_{2c}\) is given by $T_{2c}=\mathcal{J}^{(2)}/\left[2\ln\left(1+\sqrt{q}\right)-\beta_1\mathcal{J}^{(1)}\right]$ . Parameters: $\mathcal{J}^{(1)}=1$, $\mathcal{J}^{(2)}=2$ and $\beta_1=1$.}}

     \label{fig2}
\end{figure}

\begin{figure}
    \centering
     \includegraphics[width=.48\textwidth]{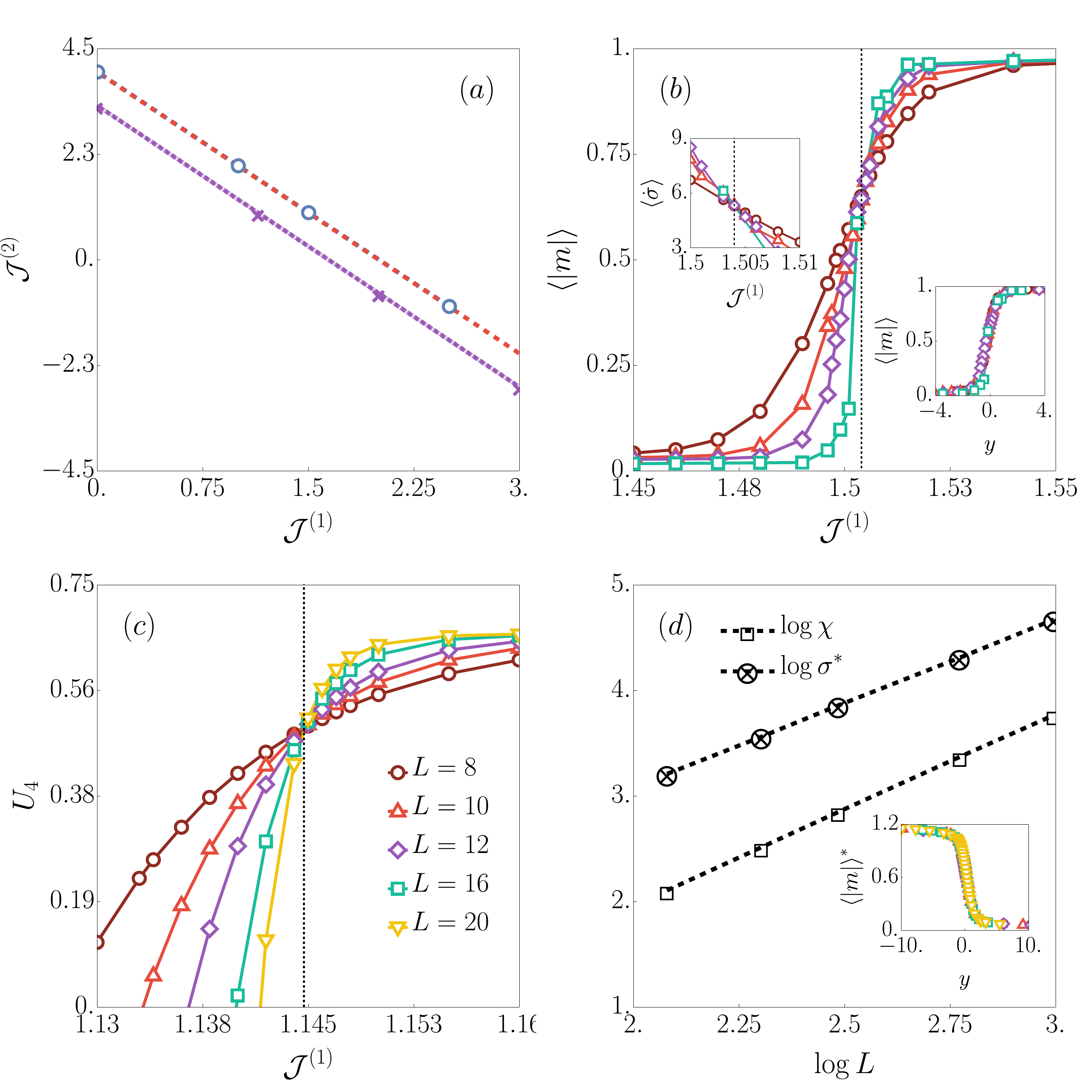}
     \caption{
     \APV{Results for the \(L\times L\) square-lattice BC model with \(\beta_1=2\), \(\beta_2=1\) and \(\Delta^{(2)}=0\). (a) Points are estimates for the location of either discontinuous or tricritical transitions. In the discontinuous case, the points follow the (dashed) line \(\frac{1}{2}\beta_1\mathcal{J}^{(1)} + \frac{1}{2}\beta_2\mathcal{J}^{(2)} = 2\), with \(\frac{1}{2}\beta_1\Delta^{(1)}+\frac{1}{2}\beta_2\Delta^{(2)}=3.984\), while in the tricritical case the (dotted) line corresponds to \(\frac{1}{2}\beta_1\mathcal{J}^{(1)} + \frac{1}{2}\beta_2\mathcal{J}^{(2)} = 1.642\), with \(\frac{1}{2}\beta_1\Delta^{(1)}+\frac{1}{2}\beta_2\Delta^{(2)}=3.233\). Both lines nicely follow the predictions of Eq.~\eqref{eqe}, according to the equilibrium data in Refs.~\cite{Silva2006, PhysRevE.92.022134}.  (b) Behavior of \(\langle|m|\rangle\) (main panel) and of \(\langle\dot\sigma\rangle\) (upper inset) as functions of \(\mathcal{J}^{(1)}\), for various values of \(L\) and the particular choice \(\left(\mathcal{J}^{(2)},\Delta^{(1)}\right)=(1,3.984)\) along the discontinuous line in (a). The lower inset shows \(\langle|m|\rangle\) as a function of the rescaled distance to the transition, measured by \(y=\left(\mathcal{J}^{(1)}-\mathcal{J}_c^{(1)}\right)L^2\), in which \(\mathcal{J}_c^{(1)}\) is the value of \(\mathcal{J}_c^{(1)}\) signaled by the vertical dotted line in the main panel, marking the discontinuous transition for the informed parameters according to Eq.~\eqref{eqe}. (c) Behavior of the Binder cumulant \(U_4\) as a function of \(\mathcal{J}^{(1)}\), for various values of \(L\) and the particular choice \(\left(\mathcal{J}^{(2)},\Delta^{(1)}\right)=(1,3.233)\) along the tricritical line in (a). The crossing of the various curves occurs at a value of \(\mathcal{J}^{(1)}\), signaled by the vertical dotted line, compatible with Eq.~\eqref{eqe}. (d) Finite-size dependence, at the same tricritical point as in (c), of \(\chi\) and of the derivative of the entropy production, \(\sigma^*\equiv d\langle {\dot \sigma} \rangle/d \mathcal{J}^{(1)}\). The corresponding slopes of the log-log plots yield estimates of the tricritical exponents  \(\gamma_t/\nu_\perp=1.82(1)\), and \(\zeta_t/\nu_\perp=1.59(1)\). Inset: rescaled order-parameter $m^*=\langle|m|\rangle L^{\beta_t/\nu_\perp}$ versus  \(y=\left(\mathcal{J}^{(1)}-\mathcal{J}_c^{(1)}\right)L^{1/\nu_\perp}\) for the equilibrium exponent \(\beta_t/\nu_\perp=3/40\). The estimate for \(\gamma_t/\nu_\perp\) is also in excellent agreement with the equilibrium
     equilibrium value \(37/20\). The value of \(\zeta/\nu_\perp\) is strongly dependent on the choice of parameters.
     }
     }
     \label{fig3}
\end{figure}

\APV{
Equation~(\ref{eqe}), together with Eq.~(\ref{eq:psinfH}), constitutes the main finding of this Letter, and some remarks about it are in order. Firstly, it is valid when a system is in fast alternating contact with two thermal baths, irrespective of the underlying equilibrium model, the nature of the phase transition (discontinuous, continuous, or even multicritical), and the lattice topology. This is illustrated in Figs.~\ref{fig1}-\ref{fig3}, which show results for various systems, detailed below. Secondly, Eq.~(\ref{eqe}) obviously reduces to the equilibrium condition when $\boldsymbol\epsilon^{(1)}=\boldsymbol\epsilon^{(2)}$ and $\beta_1=\beta_2$. Thirdly, despite the probability distribution in Eq.~(\ref{eq:psinfH}) presenting a similar form as for the equilibrium case, the dynamics
is out of equilibrium, being characterized by a positive NESS entropy production (or rather the entropy-production \emph{rate}) $\langle \dot{\sigma} \rangle$, given by \cite{broeck15}
% Voltamos mais uma vez à questão do sinal global na expressão abaixo. Acho que tem que ser como está abaixo. A propósito, deveríamos corrigir a expressão no apêndice do artigo anterior.
\begin{equation}
\langle \dot{\sigma} \rangle = \sum_{\nu} \sum_{s,s'} J_{s s^\prime}^{(\nu)} \ln \frac{\omega^{(\nu)}_{s s'}}{\omega^{(\nu)}_{s's}} =-\sum_{\nu} \beta_\nu \sum_{s,s'} J_{s s^\prime}^{(\nu)} Q_{s s^\prime}^{(\nu)}.
\label{ep}
\end{equation}
Notice that $\langle \dot{\sigma} \rangle$ can also be rewritten as $\langle \dot{\sigma} \rangle=-\sum_{\nu} \beta_\nu \langle \dot{Q}_\nu \rangle$, where \(\langle \dot{Q}_\nu \rangle\)
is the heat exchanged (per unit time) with the \( \nu \)-th thermal bath,
\begin{equation}
\langle \dot{Q}_\nu \rangle = \sum_{s,s'}J_{s s^\prime}^{(\nu)} Q_{s s^\prime}^{(\nu)}.
\label{eq:Qnu}
\end{equation}
}
%From the first law of thermodynamics $\langle \mathcal{P} \rangle + \langle %\dot{Q}_1 \rangle + \langle \dot{Q}_2 \rangle = 0$, one obtains the mean %power output \( \langle \mathcal{P} \rangle \)   given by
%\begin{equation}
%\langle \mathcal{P} \rangle = \sum_{(s,s')} \left(\Delta E^{(2)}_{s's}  - %\Delta E^{(1)}_{s's}\right)J_{s's}^{(1)},
%\end{equation}
%in such a way that all above quantities can be directly evaluated from Eqs.~%(\ref{eq:Kramers}) and (\ref{eq:psinfH}).

{\it Applications--}
\APV{
We now test our predictions with nonequilibrium versions of well-known magnetic systems, namely the Ising \cite{yeomans1992statistical}, Potts \cite{RevModPhys.54.235} and Blume-Capel \cite{Blume,Capel} models, which find wide applicability, for instance, in the description of different types of mixtures, water-like anomalies \cite{franzese, mishima1998relationship, fiore2009liquid}, and Langmuir monolayers \cite{fiorecarneiro,pelizzola2000heterochirality,RevModPhys.71.779}.
}

%The first one is the simplest system amenable exact solution and critical exponents for square lattices.  
\APV{
The spin-\(1/2\) Ising model is a paradigmatic model for continuous phase transitions, and, as already mentioned, its nonequilibrium version is characterized by parameters \(\boldsymbol{\epsilon}^{(\nu)}=\left( \mathcal{J}^{(\nu)}, h^{(\nu)} \right)\) representing the exchange constant and the magnetic field
%, associated  with a system energy \(E^{(\nu)}(s) = - \mathcal{J}^{(\nu)} \sum_{\langle i,j\rangle}s_is_j-h^{(\nu)}\sum_i s_i\) 
when in contact with the $\nu$-th thermal bath. 
From Eq.~\eqref{beh}, we have \(\tilde\epsilon_1=\frac{1}{2}\beta_1 \mathcal{J}^{(1)} + \frac{1}{2}\beta_{2} \mathcal{J}^{(2)}\) and \(\tilde\epsilon_2=\frac{1}{2}\beta_1 h^{(1)} + \frac{1}{2}\beta_{2} h^{(2)}\). The criticality conditions correspond to \(f(\boldsymbol{\tilde\epsilon})=(\tilde\epsilon_1-\phi_\mathrm{sing},\tilde\epsilon_2)=(0,0)\), with a constant \(\phi_\mathrm{sing}\) which depends on the lattice topology.
} 
%The former and latter parameters  account to  the exchange (interaction)  between neighbor spins and the magnetic field associated with $\nu$-th thermal bath, respectively. 
\APV{
We explicitly consider both the mean-field (``all-to-all'') limit, for which \(\phi_\mathrm{sing}=1\), and the square-lattice topology, for which \(\phi_\mathrm{sing}=2/\ln (1+\sqrt{2})\). We fix the values of \(\beta_1\), \(h^{(1)}\), \(\mathcal{J}^{(1)}\), and \(\mathcal{J}^{(2)}\), varying \(\beta_2\) and choosing \(h^{(2)}\) as described below. From the criticality condition \(\tilde\epsilon_1=\phi_\mathrm{sing}\), we identify a critical value \(\beta_{2c}\) of \(\beta_2\), satisfying \(\frac{1}{2}\beta_1 \mathcal{J}^{(1)} + \frac{1}{2}\beta_{2c} \mathcal{J}^{(2)} =  \phi_\mathrm{sing}\). We work with two combinations of magnetic fields compatible with  \(\tilde\epsilon_2=0\): (I) $h^{(1)}=h^{(2)}=0$; (II) $\beta_1h^{(1)}+\beta_2h^{(2)}=0$, for all values of \(\beta_2\), with \(h^{(1)}\neq0\). Notice that in combination (II) the value of \(h^{(2)}\) varies with \(\beta_2\), but the effective field \(\tilde\epsilon_2\) is always zero.
}

\APV{
In the mean-field limit, as shown in the Supplemental Material (SM),
% ACHO MELHOR DEIXARMOS ESSE DETALHE DE LADO, PARA NÃO TERMOS QUE REDISCUTIR AS CONDIÇÕES DE CRITICALIDADE ACIMA.
%which involves replacing \(\mathcal{J}^{(\nu)} \rightarrow \mathcal{J}^{(\nu)} / 2N\), 
the dynamics can be described via the occupation numbers \(N_{\pm}\) of spins in each individual state \(s\in\{+1,-1\}\). A change \(s \rightarrow -s\) corresponds to updating the occupation numbers as \(N_{\pm s} \rightarrow N_{\pm s} \mp 1\).
By taking the limit \( N\rightarrow \infty\), the dynamics is fully described  via probabilities \(p_{\pm }\rightarrow N_\pm/N\), with transition rates $\omega^{(\nu)}_ {-s,s}$ that depend on $p_\pm$ and are associated with the energy change corresponding to a single spin flipping from state \(s\) to \(-s\) in contact with bath \(\nu\). As discussed in SM, the resulting master equation can be solved via the spanning tree method \cite{schnakenberg}, leading to steady-state solutions \(p^{\mathrm{st}}_+ = (\omega^{(1)}_{+-}+\omega^{(2)}_{+-})/(\omega^{(1)}_{+-}+\omega^{(2)}_{+-}+\omega^{(1)}_{-+}+\omega^{(2)}_{-+})\) and \(p^{\mathrm{st}}_-=1-p^{\mathrm{st}}_+\), which fully agree with Eq.~(\ref{eq:psinfH}). Therefore, in terms of the magnetization $m=2p_+-1$, the steady-state solution satisfies
\begin{equation}
\tilde\epsilon_1 m+\tilde\epsilon_2 = \tanh^{-1}m.
\label{mage}
\end{equation} 
}

\APV{
In order to analyze the mean-field phase transitions, we assume that $m$ is small close to criticality, and measure the distance to criticality by using the parameter \(g\equiv\frac{1}{2}(\beta_2-\beta_{2c})\mathcal{J}^{(2)}\), in terms of which we can write \(\tilde\epsilon_1=1+g\) and \(\tilde\epsilon_2=\tilde\epsilon_2^*+g h^{(2)}/\mathcal{J}^{(2)}\), with \(\tilde\epsilon_2^*\equiv \frac{1}{2}\beta_1 h^{(1)} + \frac{1}{2}\beta_{2c} h^{(2)}\). Notice that, generically, \(\tilde\epsilon_{2}=\tilde\epsilon_2^*=0\) for field combination (I), while \(\tilde\epsilon_{2}=0\neq\tilde\epsilon_2^*\) for field combination (II). Expanding the right-hand side of Eq.~\eqref{mage} we then obtain
\begin{equation}
    \tilde\epsilon_2+g m = \frac{1}{3}m^3 + O(m^5),
\end{equation}
formally equal to the well-known expansion of the Curie--Weiss equation around the equilibrium critical point of the mean-field spin-1/2 Ising model \cite{salinas2001introduction}. 
%where  above coefficients are listed in Supplemental Material. The criticality yields for any set of $h^{(1)}$ and  $h^{(2)}$ satisfying the condition $\beta_1 h^{(1)}+\beta_2 h^{(2)}=0$ [conditions (i)-(iii)]  and $a_1=0$, implying that $\beta_1\mathcal{J}^{(2)}+\beta_1\mathcal{J}^{(2)}=2$, which is consistent with Eq.~(\ref{eqe}) and   $\phi=1$. 
Therefore, we have, for both field combinations (I) and (II), \(m\sim g^{\beta}\), for \(g\geq0\), with a critical exponent $\beta=1/2$ (not to be confused with the inverse bath temperatures), while \(m=0\) for \(g<0\).}

% SERIA MELHOR MOSTRARMOS NO MATERIAL SUPLEMENTAR O CÁLCULO DA PRODUÇÃO DE ENTROPIA NO LIMITE DE CAMPO MÉDIO. ISSO ENVOLVE REESCREVER A EQUAÇÃO GERAL DESSA GRANDEZA NA FORMA ADPTADA PARA AS POPULAÇÕES DE SPINS + E -.
\APV{
As shown in SM, we can write the mean-field entropy production as
% Substituí W por D porque W poderia ser associado a algum tipo de trabalho.
\begin{equation}
\langle{\dot \sigma}\rangle = \frac{2\left(h_\mathrm{eff}^{(1)}-h_\mathrm{eff}^{(2)}\right)\sinh\left(h_\mathrm{eff}^{(1)}-h_\mathrm{eff}^{(2)}\right)}{\cosh h_\mathrm{eff}^{(1)} - \cosh h_\mathrm{eff}^{(2)}},
\label{eq:entprod}
\end{equation}
where $h_\mathrm{eff}^{(\nu)}=\beta_1\left( h^{(\nu)}+ \mathcal{J}^{(\nu)}m\right)$. Therefore, in the paramagnetic phase, as \(m=0\), $\langle{\dot \sigma}\rangle$ is constant and equal to its value at the critical point, $\langle{\dot \sigma}\rangle_c$, which is zero under combination (I) and equals \(\langle{\dot \sigma}\rangle_c=4\beta_1 h^{(1)}\sinh\left(\beta_1 h^{(1)}\right)\) under combination (II). Note that $\langle{\dot \sigma}\rangle$ is positive (as expected) for any nonequilibrium condition but vanishes for $\beta_1 \boldsymbol\epsilon^{(1)} =\beta_2 \boldsymbol\epsilon^{(2)}$. Indeed, close to criticality, by expanding Eq.~\eqref{eq:entprod} in powers of $m$ and $g$, we see a clear distinction between the cases \(\beta_1 \mathcal{J}^{(1)}\neq \phi_\mathrm{sing}\) and \(\beta_1 \mathcal{J}^{(1)} = \phi_\mathrm{sing}\) (with \(\phi_\mathrm{sing}=1\) in the mean-field limit). This is related to the fact that \(\beta_1 \mathcal{J}^{(1)} = \phi_\mathrm{sing}\) with \(g=0\) implies that both baths independently impose a criticality condition on the system, corresponding precisely to $\beta_1 \boldsymbol\epsilon^{(1)} =\beta_2 \boldsymbol\epsilon^{(2)}$. Explicitly, denoting by \(\langle{\dot \sigma}\rangle_\mathrm{I}\) and \(\langle{\dot \sigma}\rangle_\mathrm{II}\) the entropy production under combinations (I) and (II), we have, if \(\beta_1 \mathcal{J}^{(1)}\neq \phi_\mathrm{sing}\), that \(\langle{\dot \sigma}\rangle_\mathrm{I}\sim m^2\sim g\), and \(\langle{\dot \sigma}\rangle_\mathrm{II}-\langle{\dot \sigma}\rangle_c\sim m\sim g^{1/2}\). On the other hand, if \(\beta_1 \mathcal{J}^{(1)} = \phi_\mathrm{sing}\), we have \(\langle{\dot \sigma}\rangle_\mathrm{I}\sim g^2 m^2\sim g^3\) and \(\langle{\dot \sigma}\rangle_\mathrm{II}-\langle{\dot \sigma}\rangle_c\sim m^2\sim g\). Of course, these last results are valid in the ferromagnetic phase  (\(g\gtrsim0\)). As \(\langle{\dot \sigma}\rangle\) is constant in the paramagnetic phase, it is nonanalytic at the critical point, a feature thay may manifest in the derivative \(d\langle\dot\sigma\rangle/dg\) being discontinuous or divergent. Therefore, we may formally write \(|d\langle\dot\sigma\rangle/dg|\sim g^{-\zeta}\), but with a nonuniversal critical exponent \(\zeta\), as in the case of spin systems coupled to two baths acting on different sublattices \cite{martynec2020entropy}.
}  
 
\APV{In order to analyze phase transitions in the square lattice for the same two combinations of magnetic fields, we performed numerical simulations based on the  Gillespie algorithm \cite{gillespie1977exact} and investigated thermodynamic singularities using finite-size scaling (FSS) theory \cite{landau2021guide}. Critical points were located via the crossing of curves for the Binder cumulant, defined as \(U_4=1-\langle m^4\rangle/3\langle m^2\rangle^2\), for different system sizes $L$, with $N=L^2$ being the number of spins in the lattice. Close to a transition point, the average magnetization per site, \(\langle|m|\rangle\), and its variance, \(\chi=N\left(\langle m^2\rangle-\langle|m|\rangle^2\right)\), exhibit algebraic scaling, following \(\langle|m|\rangle = L^{-\beta/\nu_\perp}\,\tilde{f}_m\left(L^{1/\nu_\perp}|g|\right)\) and \(\chi = L^{\gamma/\nu_\perp}\,\tilde{f}_\chi\left(L^{1/\nu_\perp}|g|\right)\), where \(\tilde{f}_m\) and \(\tilde{f}_\chi\) are scaling functions, the variable  \(g=(\beta_2-\beta_{2 c})/\beta_{2 c}\) measures the distance to the critical point $\beta_{2 c}$, \(\beta\)  and \(\gamma\) are critical exponents associated with \(\langle|m|\rangle\) and \(\chi\), respectively, and $\nu_\perp$ is the correlation-length critical exponent. For the Ising  universality class, these exponents are known exactly, being given by $\beta=1/8,\gamma=7/4$ and $\nu_{\perp}=1$. The plots in Fig.~\ref{fig1}(b) show that Eq.~\eqref{eqe} indeed predicts the correct position of the critical point for both field combinations. By fixing $\beta_1=1/T_1=1/2$ and $\mathcal{J}^{(1)}=\mathcal{J}^{(2)}=1$, this position corresponds to $T_{2c}=\beta_2^{-1}=2/\left[2\ln\left(1+\sqrt{2}\right)-1\right]$ (vertical lines in Fig.~\ref{fig1}), a condition that is not equivalent to $\beta_1 \boldsymbol\epsilon^{(1)} =\beta_2 \boldsymbol\epsilon^{(2)}$. The scaling behaviors for $\langle|m| \rangle$ and $\chi$ remain the same under both field combinations; see insets in Figs.~\ref{fig1}(a) and (b). 
%However, under field combination (III), although \(\chi\) still follows the same critical exponent as before, this is not true for \(\langle m\rangle\), although our data were not refined enough to confirm a behavior analogous to that of the mean-field prediction, \(|\langle m\rangle|\sim |g|^{1/\delta}\), with \(\delta=15\), as expected for the 2D Ising universality class. 
In contrast with the mean-field limit, $\langle {\dot \sigma}\rangle$ does not seem to exhibit singular behavior at criticality.}
%while, as illustrated in Fig.~\ref{fig1}(c), field combination (III) leads to scaling behavior for \(d\langle {\dot \sigma} \rangle/dT_2\) following the power law $d\langle {\dot \sigma} \rangle/dT_2 = L^{\alpha'/\nu_\perp}\,\tilde{f}_\sigma \left(L^{1/\nu_\perp}|g|\right)$,  $\tilde{f}_\sigma$ being yet another scaling function. We numerically obtained $\alpha'/\nu_\perp=1.50(5)$, in contrast to the usual scaling behavior of that quantity in other systems with "up-down" $Z_2$ symmetry \cite{noa2019,aguilera2023nonequilibrium},  which corresponds to $\alpha'/\nu_\perp=0$.}

%In order to verify  Eq.~(\ref{eqe}) in such case and its independence on $h^{(\nu)}$'s,  Fig.~\ref{fig1} depicts the behavior of $\langle |m|\rangle,\chi$ and $\langle {\dot \sigma}\rangle$ versus
%$\beta_2^{-1}$ for fixed  $\beta_1,h^{(1)}\neq 0$
%and $h^{(2)}=\beta_1h^{(1)}/\beta_{2c}$. Note that $h^{(\nu)}\neq 0$ off
%the criticality and $\beta_1h^{(1)}+\beta_{2c}h^{(2)}=0$ at %$\beta_{2c}$. The
%criticality not only follows Eq.~(\ref{eqe}) (dashed vertical line) but %also critical exponents follow Ising
%universality class (see e.g. different data collapse).

\APV{
As a second application, we investigate the nonequilibrium
features of the zero-field Potts model, with energies given by 
\begin{equation}
E^{(\nu)}(s) =  -\mathcal{J}^{(\nu)}\sum_{(i,j)} \delta_{s_i, s_j},
\end{equation}
where $s_i\in\{0,1,...,q-1\}$. Thus, the parameter vector has a single component, \(\boldsymbol{\epsilon}^{(\nu)}=\mathcal{J}^{(\nu)} \). Apart from a rescaling of the exchange constants by a factor of \(2\), the case \(q=2\) is equivalent to the zero-field Ising model, so we focus here on \(q\ge3\). At equilibrium, the model exhibits a high-temperature disordered phase and a low-temperature ordered phase in which a $q$-fold degeneracy is broken. While mean-field theory predicts a discontinuous order-disorder phase transition for $q\ge 3$, in 2D the model undergoes a  discontinuous (continuous) transition for $q>4$ ($q\le 4$). We observe the same features in the nonequilibrium case, as we now describe.
}

\APV{
The order parameter for the Potts model is defined as $m=\left(q \langle p_\mathrm{max}\rangle-1\right)/(q-1)$ \cite{RevModPhys.54.235,landau2021guide,fiore2021current,fiorejcp2013}, in which \(p_\mathrm{max}=\max\{p_i\}\), \(p_i\) being the steady-state fraction of spins in the state \(s_i\). In the mean-field limit (all-to-all interactions), it is natural to formulate the nonequilibrium dynamics in terms of the set \(\{p_i\}\), as described in SM. For any value of \(q\), the NESS order parameter \(m\) can be shown to satisfy the self-consistency equation
\begin{equation}
    \frac{1}{2} \left(\beta _1  \mathcal{J}^{(1)} +\beta _2  \mathcal{J}^{(2)}\right)m = \ln\frac{1+(q-1)m}{1-m}.
   \label{PottsMinimum}
\end{equation}
 Notice that Eq.~\eqref{PottsMinimum}, for which \(m=0\) is always a solution, is the nonequilibrium analog of the corresponding equilibrium expression for the mean-field Potts model, obtained from the minimization of a free-energy functional \cite{RevModPhys.54.235}.
For \(q\ge3\), with fixed values of \(\beta_1\), \(\mathcal{J}^{(1)}\), and \(\mathcal{J}^{(2)}\), there is a range of values of \(\beta_2\) for which Eq.~\eqref{PottsMinimum} has two positive solutions for \(m\), signaling the existence of a spinodal region, characteristic of a discontinuous phase transition. Within that range, we numerically checked by integrating the dynamical equations that the disordered (\(m=0\)) solution and the largest positive solution are always stable to small perturbations, while the smallest positive solution is always unstable. By combining Eq.~\eqref{eqe} with the equilibrium condition for phase coexistence \cite{RevModPhys.54.235}, we obtain the presumed nonequilibrium condition 
\begin{equation}
\frac{1}{2}\left(\beta_1\mathcal{J}^{(1)}+\beta_2\mathcal{J}^{(2)}\right)=\frac{2(q-1)}{q-2}\ln (q-1).
\label{eq:pottscrit}
\end{equation}
%In the limit \(q\rightarrow 2^+\), and apart from a rescaling of the exchange coupling by a factor of \(2\), Eq.~\eqref{eq:pottscrit} reduces to that for the mean-field Ising model, previously discusssed. 
For \(q\ge3\), we numerically checked that the value of \(\beta_2\) obtained from Eq.~\eqref{eq:pottscrit} always lies within the range of values associated with the simultaneous stability of the ordered and the disordered solutions for \(m\).
}

\APV{
The mean-field NESS entropy production $\mom{\dot{\sigma}}$ is given by
\begin{equation}
    \mom{\dot{\sigma}}=\frac{4(q-1)\left(K_\mathbf{eff}^{(1)}-K_\mathbf{eff}^{(2)}\right)\sinh\left(K_\mathbf{eff}^{(1)}-K_\mathbf{eff}^{(2)}\right)}{q e^{K_\mathbf{eff}^{(1)}+K_\mathbf{eff}^{(2)}}-2\sinh\left(K_\mathbf{eff}^{(1)}+K_\mathbf{eff}^{(2)}\right)},
    %\mom{\dot{\sigma}}=\frac{ (q-1) m\left(\beta _1 \mathcal{J}^{(1)}-\beta _2 \mathcal{J}^{(2)}\right) \sinh \left[\frac{1}{4} m \left(\beta _1 \mathcal{J}^{(1)}-\beta _2 \mathcal{J}^{(2)}\right)\right]}{q e^{\frac{1}{4} m \left(\beta _1 \mathcal{J}^{(1)}+\beta _2 \mathcal{J}^{(2)}\right)}-2 \sinh \left[\frac{1}{4} m \left(\beta _1 \mathcal{J}^{(1)}+\beta _2 \mathcal{J}^{(2)}\right)\right]}.
\end{equation}
with \(K_\mathrm{eff}^{(\nu)}=\frac{1}{4}\beta_\nu\mathcal{J}^{(\nu)}m\). As in the case of the Ising model with field combination (I), \(\mom{\dot{\sigma}}\) is zero in the disordered phase. For \(q\ge3\) and \(\beta_1\mathcal{J}^{(1)}\neq\beta_2\mathcal{J}^{(2)}\) at the transition point, \(\mom{\dot{\sigma}}\) has a discontinuity at the transition, while if  \(\beta_1\mathcal{J}^{(1)}=\beta_2\mathcal{J}^{(2)}\) at the transition point, \(\mom{\dot{\sigma}}\) is nonzero in the ordered phase but approaches zero continuously at the transition.
}

\APV{
For the square-lattice Potts model, based on the equilibrium condition for the phase-transition point \cite{Beffara2011}, Eq.~\eqref{eqe} predicts a transition line given by
\begin{equation}
\frac{1}{2}\left(\beta_1\mathcal{J}^{(1)}+\beta_2\mathcal{J}^{(2)}\right)=\ln\left(1 + \sqrt{q}\right),    
\end{equation}
for all values of $q$. We confirmed this prediction via numerical simulations of \(L\times L\) lattices using the Gillespie algorithm \cite{gillespie1977exact}. For \(q=3\)  (see Fig.~\ref{fig2}a), we obtained scaling behaviors of \(\langle m\rangle\) and its susceptibility-like variance \(\chi\equiv L^2\left(\mom{m^2}-\mom{m}^2\right)\) that are consistent with the exact (equilibrium) critical exponents $\beta=1/9$, $\gamma=13/9$, and $\nu_{\perp}=5/6$. On the other hand, for \(q>4\), and in agreement with the expectation of a discontinuous phase transition, we observed that, as in Refs.~\cite{fss, fss2, noa2019, goes}, \(\chi\) scales at the transition point as \(\chi\sim L^2\)
%, while the curves for \(\langle m \rangle\) with different system sizes cross at the transition point 
(see Fig.~\ref{fig2}b). 
As in the case of the Ising model (\(q=2\)), $\langle {\dot \sigma}\rangle$ is analytic for $q=3$ (not shown), while for \(q>4\) there is a clear distinction between the finite-size behavior in the ordered and in the disordered phase, possibly with a discontinuity at the transition point. The precise characterization of this behavior will be the subject of a future investigation.
}

\APV{As a final application, we look at the Blume-Capel (BC) model \cite{Blume,Capel}, a particular case of the Blume-Emery-Griffiths model \cite{beg,PhysRevLett.67.1027}, which is a paradigmatic spin-1 Ising system displaying tricritical behavior. The issue of nonequilibrium tricriticality has attracted recent interest in both classical  \cite{PhysRevLett.131.017102} and quantum systems \cite{friedemann2018quantum,PhysRevLett.133.223401}. Nevertheless, little is know about the scaling behavior of genuine nonequilibrium quantities (e.g. entropy production) at tricriticality.
}

\APV{
We investigated a nonequilibrium BC model characterized, at zero magnetic field, by the parameter vector \(\boldsymbol{\epsilon}^{(\nu)}=\left( \mathcal{J}^{(\nu)},\Delta^{(\nu)} \right)\) with total energy
\begin{equation}
E^{(\nu)}(s) =  -\mathcal{J}^{(\nu)}\sum_{(i,j)} s_i s_j + 
%h^{(\nu)}\sum_{s_i}s_i+
\Delta^{(\nu)} \sum_{i}s_i^2,
\end{equation}
where  $s_i\in\{0,\pm 1\}$ and \(\mathcal{J}^{(\nu)}>0\), while the crystal-field parameters $\Delta^{(\nu)}$ add a constant energy contribution for each spin in a state $s\neq 0$.
In the equilibrium version, corresponding to \(\boldsymbol{\epsilon}^{(1)} = \boldsymbol{\epsilon}^{(2)}\equiv(\mathcal{J},\Delta)\) and \(\beta_1=\beta_2\equiv1/T\), there is a tricritical point \(P_t\) separating regions of discontinuous and of continuous transitions, and whose location in the \((\Delta/\mathcal{J})\times (T/\mathcal{J})\) plane depends on the lattice topology. In the all-to-all version, \(P_t\) has coordinates \((T_t/\mathcal{J},\Delta_t/\mathcal{J})=(\frac{1}{3},\frac{1}{3}\ln4)\) \cite{Capel}, while in the square lattice \((T_t/\mathcal{J}, \Delta_t/\mathcal{J}) \approx (0.608, 1.966)\) \cite{Silva2006,PhysRevE.92.022134}. 
}

\APV{
As with previous models, we first consider the all-to-all limit. The treatment described in SM leads to the conclusion that the nature of the phase transition depends on two coefficients of an expansion of the equation giving the steady-state solution for the magnetization \(m\), namely
\begin{eqnarray}
a_1&\propto&  \frac{1}{2}\left(\beta_1\mathcal{J}^{(1)}+\beta_{2}\mathcal{J}^{(2)}\right)-\left[1+\frac{1}{2}e^{\frac{1}{2}\left(\beta_1 \Delta^{(1)}+\beta_2 \Delta^{(2)}\right)}\right],\\
a_3&\propto&4-e^{\frac{1}{2}\left(\beta_1 \Delta^{(1)}+\beta_2 \Delta^{(2)}\right)}.
\end{eqnarray}
}

\APV{
A simple critical point occurs for \(a_1=0\), \(a_3<0\), so that the criticality condition clearly reduces to the equilibrium one, \( \beta \mathcal{J} = 1+ \frac{1}{2} e^{\beta\Delta}\) \cite{Capel}, when the two reservoirs have the same parameters, in agreement with Eq.~\eqref{eqe}. If we deviate from the criticality condition by changing one of the parameters by a small amount \(g\), then, in the ordered phase, the magnetization scales as \(m\sim |g|^{\beta_c}\), with a critical exponent \(\beta_c=1/2\) (see SM). 
}

\APV{
A tricritical point corresponds to \(a_1=a_3=0\), leading to 
\begin{eqnarray}
    \frac{1}{2}\left(\beta_1\mathcal{J}^{(1)}+\beta_{2}\mathcal{J}^{(2)}\right) &=& 3, \\
    \frac{1}{2}\left(\beta_1 \Delta^{(1)}+\beta_2 \Delta^{(2)}\right) &=& \ln 4,
\end{eqnarray}
again in agreement with Eq.~\eqref{eqe}. A small deviation \(g\) from the tricriticality condition yields a magnetization scaling as \(m\sim |g|^{\beta_t})\), with a tricritical exponent \(\beta_t=1/4\) (see SM).
}

\APV{
In the paramagnetic phase, the mean-field entropy production \(\langle\dot\sigma\rangle\) depends on the products \(\beta_\nu \Delta^{(\nu)}\), being given by
\begin{equation}
\langle\dot\sigma\rangle_\mathrm{para} = \frac{4\left(e^{\frac{1}{2}\beta_1 \Delta^{(1)}}-e^{\frac{1}{2}\beta_2 \Delta^{(2)}}\right)}{2+e^{\frac{1}{2}\left(\beta_1 \Delta^{(1)}+\beta_2 \Delta^{(2)}\right)}}\left(\beta_1\Delta^{(1)}-\beta_2\Delta^{(2)}\right).
\end{equation}
As discussed in SM, the scaling of \(\langle\dot\sigma\rangle\) in the ferromagnetic phase close to a critical (\(i=c\)) or tricritical (\(i=t)\) point follows
\begin{equation}
\langle\dot\sigma\rangle-\langle\dot\sigma\rangle_c\sim|g|^{1-\zeta_i},
\end{equation}
with \( \zeta_t = 1/2 \) at a tricritical point. On the other hand, \( \zeta_c = 0 \) at a critical point, as in the case of the Ising model under field combination (I) and of the systems investigated in Ref.~\cite{noa2019}. This is valid for a ``generic'' criticality conditions, i.e. one at which \(\beta_1\mathcal{J}^{(1)}\neq \beta_2\mathcal{J}^{(2)}\) and \(\beta_1\Delta^{(1)}\neq \beta_2\Delta^{(2)}\). In contrast, under an ``independent'' criticality condition, at which \(\beta_1\mathcal{J}^{(1)}= \beta_2\mathcal{J}^{(2)}\) and \(\beta_1\Delta^{(1)}=\beta_2\Delta^{(2)}\), it can be checked that \(\zeta_c=4\) and \(\zeta_t=1/2\). 
}
%Likewise, the expansion of both \( \langle {\cal P} \rangle \) and \( \langle{\dot %Q}_\nu\rangle \) in powers of \( m \) leads to
%the same scaling behavior to $\langle {\dot \sigma}\rangle$ given by 
%$\langle{\dot Q}_{\nu'}\rangle \sim \langle{\dot Q}_{\nu'}\rangle_{dis} + d_{q\nu'}%(X_{\nu i}-X_\nu)^{1-\alpha_i}$ 
%and  
%$\langle{\cal P}\rangle \sim -\sum_{\nu'}\left(\langle{\dot Q}_{\nu'}\rangle_{dis} %+ d_{q\nu'}(X_{\nu i}-X_{\nu})^{1-\alpha_i}\right)$, respectively, 
%where \( \langle{\dot Q}_{\nu'}\rangle_{dis} = 2\Delta_{\nu'}(e^{\frac{\beta_1 \Delta^{(1)}}{2}} - e^{\frac{\beta_2 \Delta^{(2)}}{2}})/a_2 \) and \( d_{q\nu'} = A^2_i b_{q\nu'} \) (see Supplemental Material). 

\APV{
Finally, the main properties of the nonequilibrium \(L\times L\) square-lattice BC model are shown Fig.~\ref{fig3}. We fix the values of \(\beta_1\), \(\beta_2\), \(\Delta_1\), \(\Delta_2\), and \(\mathcal{J}^{(2)}\), varying \(\mathcal{J}^{(1)}\) in search of order-disorder transitions. Discontinuous transitions are located by the crossing of \(\langle |m|\rangle\) or $\langle {\dot \sigma}\rangle$ curves for different values of \(L\) [Fig.~\ref{fig3}(b)], while the location of continuous transitions is estimated from the crossing of curves for the Binder cumulant [Fig.~\ref{fig3}(c)]. As shown in Fig.~\ref{fig3}(a), the estimated transition points follow Eq.~(\ref{eqe}), with the equilibrium transition conditions extracted from Refs.~\cite{Silva2006,PhysRevE.92.022134}. The values obtained for the nonequilibrium tricritical exponents are compatible with the equilibrium ones, given by \(\beta_t/\nu_\perp = 3/40\) and \(\gamma_t/\nu_\perp = 37/20\)~\cite{moueddene2024critical} for for $\langle |m|\rangle$ and $\chi$, respectively [Fig.~\ref{fig3}(d)].
}
%, whose discontinuous phase transitions is characterized by the crossing among $\langle |m|\rangle$ [Fig.~\ref{fig3}b] and $\langle {\dot \sigma}\rangle$ [upper inset] curves for different system sizes. The tricriticality is signed by entirely different set of critical exponents, consistent with  values \(\beta_t/\nu_\perp = 0.0453(2)\)~\cite{drugo}, \(\gamma/\nu_\perp = 37/20\)~\cite{moueddene2024critical}, for $\langle |m|\rangle$ and $\chi$, respectively.
%\alpha/\nu_\perp=6/5$ for the
%order-parameter, susceptibility and specific heat, respectively, such latter ones 
%very recently obtained at Ref.~\cite{moueddene2024critical}. 
%Fig.~\ref{fig3}
%summarizes our main findings for both all-to-all and square-lattice
%interactions for a discontinuous transition point and at the tricriticality. 

\APV{
 While the behavior of the entropy production
  at a critical point (not shown) is akin
  to previous examples for two-dimensional systems, it is remarkably different at a tricritical point [see e.g. Fig.~\ref{fig3}(c)]. The derivative $\sigma^*\equiv d\langle {\dot \sigma} \rangle/d g$ scales as \(L^{-\zeta/\nu_\perp}\), but the exponent is nonuniversal. This is similar to what is observed in an investigation of continuous transitions in a square-lattice version of the four-state clock model \cite{martynec2020entropy} subject to two baths with distinct temperatures acting on spins in different sublattices. The precise characterization of the nonuniversal behavior of the tricritical entropy production will be the subject of future investigations.
}

{\it Conclusions--}
\APV{
In this work, we have derived an exact mapping between the nonequilibrium steady states of systems in fast alternating contact with two thermal baths and the equilibrium distributions of suitably defined models. This mapping exactly determines the location of nonequilibrium phase transitions, irrespective of model details, interaction topology, or type of transition, and we confirmed its validity in Ising, Potts, and Blume-Capel systems. As detailed in the Supplemental Material, the mapping  remains approximately valid when the alternation rate \(d\) is finite, with corrections of order \(1/d\).}

\APV{
Our mapping provides a predictive tool, allowing knowledge on equilibrium critical phenomena to be directly applied to a range of nonequilibrium settings. Our results hint at applications in stochastic thermodynamics and microscopic thermal machines, while also suggesting that experimental signatures of criticality in systems with thermal gradients, for instance in quantum dots or magnetic junctions, could be interpreted through the simpler lens of equilibrium phase transitions. 
%In summary, we derived an exact mapping that links the probability distributions and phase-transition conditions of nonequilibrium systems in contact with two thermal baths to those of corresponding equilibrium models. In the limit of fast alternating contact with the baths, this result holds irrespective of model details, interaction topology, or type of transition, and we confirmed its validity in Ising, Potts, and Blume-Capel systems. Furthermore, we gave evidence that the result remains approximately valid when the alternation rate is finite.
}
 %Our work paves the way for deeply understanding the key aspects about probability distribution and phase transitions out of equilibrium. Exact probability distributions of systems simultaneously placed in contact with two thermal baths were obtained, irrespective the model details and  lattice topologies. As examples of our discoveries,  nonequilibrium couterparts of three icon equilibrium statistical mechanic models were investigated, namely Ising, Potts and Blume-Emery-Griffiths. As an important were obtained. As a final remark, it is important to note that, despite the generality of the probability distribution valid for simultaneous contact, results all previous examples involving non-simultaneous thermal baths—where a switching rate $d$ between thermal baths is introduced—demonstrate that the deviation between transition points scales as $1/d$. Consequently, as $d$ increases, the probability distribution converges to the form given in Eq.~(\ref{eq:psinfH}).

{\it Acknowledgments--}\APV{
We acknowledge the financial support from Brazilian agencies CNPq and FAPESP under grants 2022/15453-0, 2023/12490-4,  2022/16192-5, 2023/17704-2,  and 2024/03763-0, respectively.
}

\bibliography{refs}

\newpage
\clearpage

\appendix            

\onecolumngrid
\begin{center}
\textbf{\large Supplemental Material: Exact Mapping of Nonequilibrium to Equilibrium Phase Transitions of Systems in Contact with Two Thermal Baths}
\end{center}

\APV{
This supplemental material is structured as follows: Sec.~\ref{apa} presents the models and the main expressions for their all-to-all descriptions. Sec.~\ref{apb} exemplifies the probability distributions for the Ising model for both finite N and in the thermodynamic limit ($N\rightarrow \infty$), along with the main corresponding thermodynamic quantities. In Sec.~\ref{d}, we show the convergence to the probability distribution given by Eq.~(\ref{eq:psinfH}) as the contact-alternation rate $d$ increases. Finally, Secs.~\ref{apd} and \ref{apc} extend this analysis to the Potts and Blume-Capel models, respectively, evaluating their thermodynamic quantities using both formulations [Eqs.~(\ref{eq:pJJ}) and (\ref{eqe})].
}
%This supplemental material is structured as follows:  In Sec.~\ref{apa},  models and the main expressions for their all-to-all descriptions were presented. Secs~\ref{apb} exemplifies the probability distributions for the Ising model for $N$ finite and $N\rightarrow \infty$ as well their main for thermodynamic quantities via both formulations. In \ref{d}, the convergence to the probability distribution given by \ref{eq:psinfH} as $d$ increases. Secs.~\ref{apc} and \ref{apd} extends the evaluation of thermodynamic quantities via both formulations [Eqs.~(\ref{eq:pJJ}) and (\ref{eqe})] for the Potts and Blume-Capel models, respectively.

\section{Models in the all-to-all formulation}\label{apa}

%Our starting point is a collection of units, to each one is attached 
%the spin variable $s_i=0,\pm $, and
%the system energy given  for $E^{(\nu)}(s)$ given by the following expression

\APV{
For all-to-all interactions, the energy expressions for the Ising, Potts, and Blume-Capel (BC) models are given by
\begin{equation}
E^{(\nu)}(s) = -\frac{\mathcal{J}^{(\nu)}}{2N}\left(\sum_{s_i}s_i\right)^2 - h^{(\nu)}\sum_{s_i}s_i,
\label{isingall}
\end{equation}
\begin{equation}
E^{(\nu)}(s) = -\frac{\mathcal{J}^{(\nu)}}{2N} \sum_{(i,j)}\delta_{s_i,s_j},
\label{pottsall}
\end{equation}
and
\begin{equation}
E^{(\nu)}(s) = -\frac{\mathcal{J}^{(\nu)}}{2N}\left(\sum_{s_i}s_i\right)^2 - h^{(\nu)}\sum_{s_i}s_i + \Delta^{(\nu)}\sum_{s_i}s_i^2,
\label{bcall}
\end{equation}
respectively, where the spin variables take the values
$s_i = \pm 1$ (Ising), $s_i = 0,\ldots,q-1$ (Potts), and $s_i = 0,\pm 1$ (BC). The division of the coupling constants by \(N\) is required to ensure the existence of the thermodynamic limit.
}

\APV{
All these energies can be alternatively expressed in terms of the total population of each state as
\begin{equation}
E^{(\nu)}(N_+,N_-,N) = -\frac{\mathcal{J}^{(\nu)}}{2N}\left[ N_+(N_+-1) + N_-(N_--1) - 2N_+N_- \right] - h^{(\nu)}(N_+ - N_-),
\label{ising}
\end{equation}
for the Ising model ($ N = N_+ + N_- $),
\begin{equation}
E^{(\nu)}(N_0,N_1,\ldots,N_{q-1},N) = -\frac{\mathcal{J}^{(\nu)}}{2N} \sum_{i=0}^{q-1} N_i(N_i-1),
\label{potts}
\end{equation}
for the Potts model ($\sum_{i=0}^{q-1} N_i = N$), and
\begin{equation}
E^{(\nu)}(N_+,N_-,N) = -\frac{\mathcal{J}^{(\nu)}}{2N}\left[ N_+(N_+-1) + N_-(N_--1) - 2N_+N_- \right] - h^{(\nu)}(N_+ - N_-) + \Delta^{(\nu)}(N_+ + N_-),
\label{eq:BCmf}
\end{equation}
for the BC model ($N_0 = N - N_+ - N_-$).
}

\APV{
A single spin flipping from an allowed state \(s_j\) to a different state \(s_i\) slightly modifies the populations and leads to an energy change \(\Delta E_{s_i,s_j}\), with the property that \(\Delta E_{s_j,s_i}=-\Delta E_{s_i,s_j}\). Explicitly, these energy changes are given by
\begin{equation}
\label{eqqq}
\Delta E^{(\nu)}_{+-} = E^{(\nu)}(N_+ + 1,N_- -1,N) - E^{(\nu)}(N_+,N_-,N) = -\frac{2\mathcal{J}^{(\nu)}}{N}(2N_+ - N + 1) - 2h^{(\nu)},
\end{equation}
for the Ising model,
\begin{equation}
\Delta E^{(\nu)}_{ij} = E^{(\nu)}(N_0,\ldots,N_i +1,\ldots,N_j-1,\ldots ,N_{q-1},N) - E^{(\nu)}(N_0,\ldots,N_i,\ldots,N_j,\ldots ,N_{q-1},N) = -\frac{\mathcal{J}^{(\nu)}}{N}(1 + N_i - N_j),
\label{eq:DeltaEPotts}
\end{equation}
for the Potts model, while for the BC model we have
\begin{eqnarray}
\Delta E^{(\nu)}_{+-} &=& E^{(\nu)}(N_+ +1,N_- -1,N) - E^{(\nu)}(N_+,N_-,N) = -\frac{2\mathcal{J}^{(\nu)}}{N}(N_+ - N_- + 1) - 2h^{(\nu)}, \nonumber\\
\Delta E^{(\nu)}_{+0} &=& E^{(\nu)}(N_+ +1,N_-,N) - E^{(\nu)}(N_+,N_-,N) =-\frac{\mathcal{J}^{(\nu)}}{N}(N_+ - N_-) - h^{(\nu)} + \Delta^{(\nu)}, \nonumber\\
\Delta E^{(\nu)}_{0-} &=& E^{(\nu)}(N_+,N_- -1,N) - E^{(\nu)}(N_+,N_-,N) = -\frac{\mathcal{J}^{(\nu)}}{N}(N_+ - N_-) - h^{(\nu)} - \Delta^{(\nu)}.
\end{eqnarray}
}

\APV{
In the thermodynamic limit, $N \rightarrow \infty$ with $N_{i}/N \equiv p_i$, the energy changes become
\begin{equation}
\Delta E^{(\nu)}_{+-} = -2\mathcal{J}^{(\nu)}(2p_+ - 1) - 2h^{(\nu)},
\end{equation}
and
\begin{equation}
\Delta E^{(\nu)}_{ij} = -\mathcal{J}^{(\nu)}(p_i - p_j),
\end{equation}
for the Ising and Potts models, respectively, and
\begin{eqnarray}
\label{eqqq2}
\Delta E^{(\nu)}_{+-} &=& -2\mathcal{J}^{(\nu)}(p_+ - p_-) - 2h^{(\nu)}, \nonumber\\
\Delta E^{(\nu)}_{+0} &=& -\mathcal{J}^{(\nu)}(p_+ - p_-) - h^{(\nu)} + \Delta^{(\nu)}, \nonumber\\
\Delta E^{(\nu)}_{0-} &=& -\mathcal{J}^{(\nu)}(p_+ - p_-) - h^{(\nu)} - \Delta^{(\nu)},
\label{eq:BCmfDE}
\end{eqnarray}
for the BC model.
}

\section{Nonequilibrium Ising model and the steady-state solution for all-to-all interactions}\label{apb}

\APV{
To illustrate that Eq.~(\ref{eq:psinfH}) is indeed a solution to Eq.~(\ref{eq:pAB}), we consider the all-to-all Ising model for both finite system sizes and the thermodynamic limit $N \rightarrow \infty$. The energy of a configuration is given by Eq.~(\ref{ising}), when the system is in contact with bath \(\nu\). 
}

\APV{
For finite $N$,  all possible transitions can be enumerated via the spanning-tree method~\cite{schnakenberg,gatien,mamede2023}. In this case, the steady-state probability $p^{\mathrm{st}}(N_+,N_-,N)$ that a system with \(N\) spins has \(N_+\) spins \(+1\) and \(N_-=N-N_+\) spins \(-1\) is given by
\begin{equation}
p^{\mathrm{st}}(N_+,N_-,N) = 
\frac{1}{Z}
\left[\prod_{n=0}^{N_+-1}\!\left(\Omega^{(1)}_{n+1,n} + \Omega^{(2)}_{n+1,n}\right)\right]
\left[\prod_{n=N_+ +1}^{N}\!\left(\Omega^{(1)}_{n-1,n} + \Omega^{(2)}_{n-1,n}\right)\right],
\label{alltoall}
\end{equation}
where $Z$ is a normalization factor,
$\Omega^{(\nu)}_{n+1,n} = 2 (N-n)\, e^{-\frac{\beta_\nu}{2}\Delta E^{(\nu)}_{+-}}$ and  
$\Omega^{(\nu)}_{n-1,n} = 2 n\, e^{-\frac{\beta_\nu}{2}\Delta E^{(\nu)}_{-+}}$,  
with the energy change $\Delta E^{(\nu)}_{+-}$ given by Eq.~(\ref{eqqq}), and \(\Delta E^{(\nu)}_{-+}=-\Delta E^{(\nu)}_{+-}\).
}

\begin{figure}[h!]
\centering
\includegraphics[width=.8\textwidth]{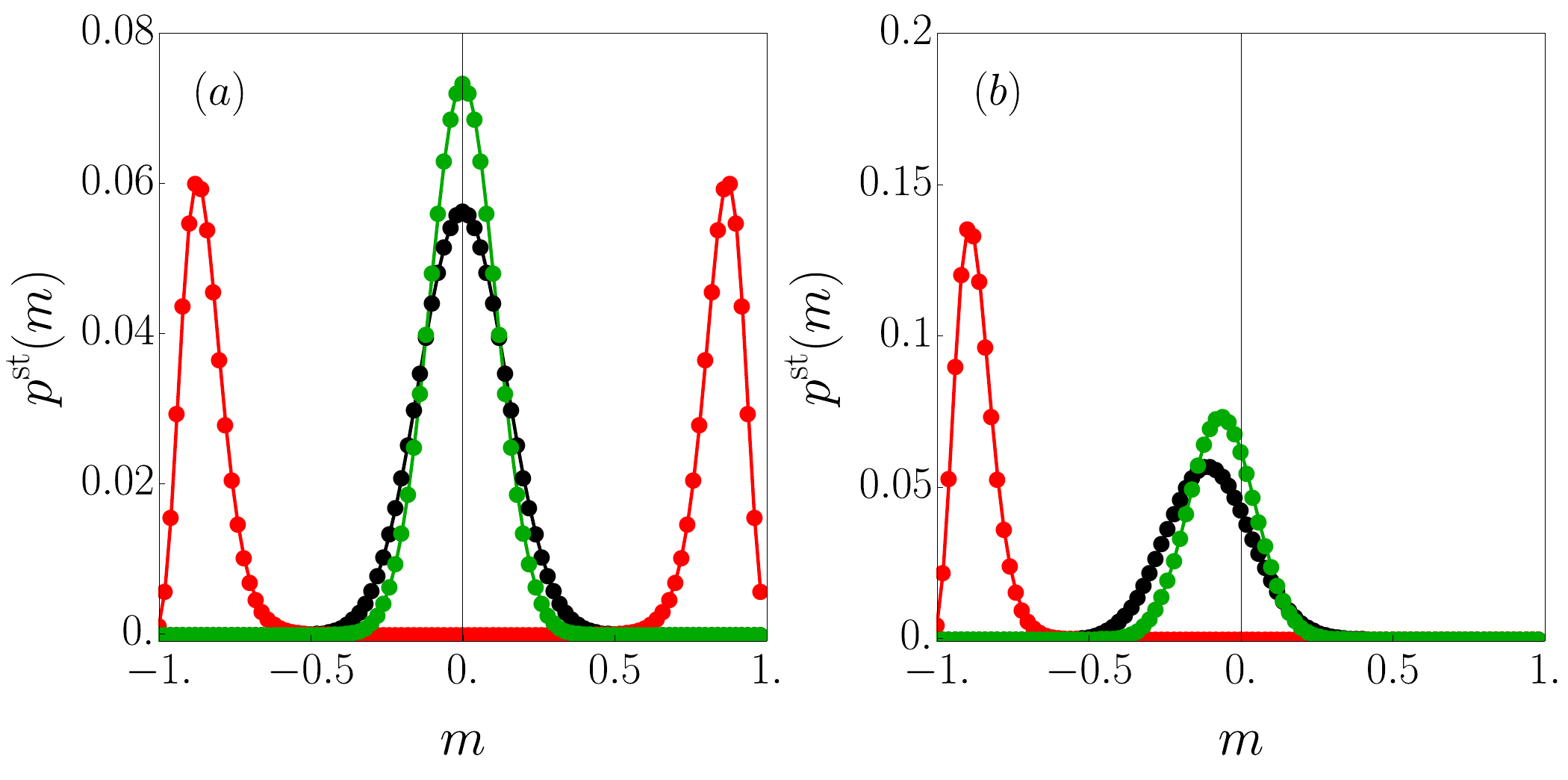}
\caption{
\APV{
Comparison between steady-state order-parameter probability distributions $p^{\mathrm{st}}(m)\equiv p^\mathrm{st}\left((1+m)N/2,(1-m)N/2,N\right)$ obtained from Eq.~(\ref{eq:pstexp}) (solid lines) and Eq.~(\ref{alltoall}) (symbols).  
Left panel corresponds to $h^{(1)}=1/4$, $h^{(2)}=-1/2$, while right panel corresponds to $h^{(1)}=1/8$, $h^{(2)}=-1/7$.  
Black curves: $\mathcal{J}^{(1)}=\mathcal{J}^{(2)}=1/4$. Red curves: $\mathcal{J}^{(1)}=\mathcal{J}^{(2)}=1/2$. Green curves: $\mathcal{J}^{(1)}=\mathcal{J}^{(2)}=1/10$. In all cases $N=100$. }
}
\label{example}
\end{figure}

\APV{
On the other hand, the steady-state distribution obtained from Eq.~(\ref{eq:psinfH}) can be written as
\begin{equation}
\label{eq:pstexp}
p^{\mathrm{st}}(N_+,N_-,N) =
\frac{1}{Z_{12}}\,
\frac{N!}{N_+!\,N_-!}\,
\exp\left[
-\frac{1}{2}\beta_1 E^{(1)}(N_+,N_-,N) -\frac{1}{2} \beta_2 E^{(2)}(N_+,N_-,N)
\right],
\end{equation}
where $E^{(\nu)}(N_+,N_-,N)$ is defined in Eq.~(\ref{ising}).  
The equivalence between Eqs.~(\ref{eq:pstexp}) and (\ref{alltoall}) is illustrated in Fig.~\ref{example} for $N=100$ and different parameter sets. As show in the figure, the results are identical.
}

\APV{
In the thermodynamic limit ($N \rightarrow \infty$), defining \(p_\pm=N_\pm/N\), following the approach in Ref.~\cite{gatien}, it can be shown from Eq.~\eqref{eq:pAB} that these quantities satisfy
\begin{eqnarray}
\dot{p}_+ &=& \left( \omega^{(1)}_{+-} + \omega^{(2)}_{+-} \right) p_-(t) - \left( \omega^{(1)}_{-+} + \omega^{(2)}_{-+} \right) p_+(t),\nonumber\\
\dot{p}_- &=& \left( \omega^{(1)}_{-+} + \omega^{(2)}_{-+} \right) p_+(t) - \left( \omega^{(1)}_{+-} + \omega^{(2)}_{+-} \right) p_-(t),\nonumber
\end{eqnarray}
with 
\[
\omega^{(\nu)}_{\pm\mp} = e^{-\frac{1}{2}\beta_\nu\Delta E_{\pm\mp}} = \exp\left\{\pm\beta_\nu\left[\mathcal{J}^{(\nu)}\left(2p^{\mathrm{st}}_+-1\right) + h^{(\nu)}\right]\right\},
\]
so that the steady-state condition is given by  
\[
(\omega^{(1)}_{-+}+\omega^{(2)}_{-+})p^{\mathrm{st}}_+ - (\omega^{(1)}_{+-}+\omega^{(2)}_{+-})p^{\mathrm{st}}_- = 0.
\]
Taking into account that \(p^\mathrm{st}_++p^\mathrm{st}_-=1\) and that the steady-state magnetization is \(m=p^\mathrm{st}_+-p^\mathrm{st}_-\), we obtain
\[
m = \frac{
\sinh(\beta_1 \mathcal{J}^{(1)}m + \beta_1 h^{(1)}) +
\sinh(\beta_2 \mathcal{J}^{(2)}m + \beta_2 h^{(2)})
}{
\cosh(\beta_1 \mathcal{J}^{(1)}m + \beta_1 h^{(1)}) +
\cosh(\beta_2 \mathcal{J}^{(2)}m + \beta_2 h^{(2)})
},
\]
which can be rewritten as
\begin{equation}
m = \tanh\!\left[
\frac{1}{2}\left(\beta_1 \mathcal{J}^{(1)}+\beta_2 \mathcal{J}^{(2)}\right)m +
\frac{1}{2}\left(\beta_1 h^{(1)}+\beta_2 h^{(2)}\right)
\right].
\label{mage_app}
\end{equation}
This last equation is equivalent to Eq.~\eqref{mage}.
}

\APV{
To characterize the critical point, we expand Eq.~(\ref{mage_app}) for small $m$ close to the transition, obtaining
\begin{equation}
0 = a_0 + a_1 m + a_2 m^2 + a_3 m^3 + \ldots,
\end{equation}
where
\begin{eqnarray}
a_0 &=& \tanh\!\left[\frac{1}{2}\left(\beta_1 h^{(1)}+\beta_2 h^{(2)}\right)\right],\nonumber\\
a_1 &=& \frac{\beta_1 \mathcal{J}^{(1)}+\beta_2 \mathcal{J}^{(2)}}{\cosh\left(\beta_1 h^{(1)}+\beta_2 h^{(2)}\right)+1} - 1,\nonumber\\
a_2 &=& -2\left(\beta_1 \mathcal{J}^{(1)}+\beta_2 \mathcal{J}^{(2)}\right)^2
\sinh^4\left[\frac{1}{2}\left(\beta_1 h^{(1)}+\beta_2 h^{(2)}\right)\right]
\csch^3\left(\beta_1 h^{(1)}+\beta_2 h^{(2)}\right),\nonumber\\
a_3 &=& \frac{1}{24}\left(\beta_1 \mathcal{J}^{(1)}+\beta_2 \mathcal{J}^{(2)}\right)^3
\left[\cosh\left(\beta_1 h^{(1)}+\beta_2 h^{(2)}\right) - 2\right]
\sech^4\left[\frac{1}{2}\left(\beta_1 h^{(1)}+\beta_2 h^{(2)}\right)\right].\nonumber
\end{eqnarray}
}

\APV{
Criticality occurs when $a_0 = a_2 = 0$ and $a_1 = 0$, yielding
\begin{equation}
\frac{1}{2}\left(\beta_1 h^{(1)}+\beta_2 h^{(2)}\right) = 0\quad\text{and}\quad  
\frac{1}{2}\left(\beta_1 \mathcal{J}^{(1)}+\beta_2 \mathcal{J}^{(2)}\right) = 1,  
\end{equation}
in agreement with Eq.~\eqref{eqe} in the main text.  
At zero effective field, and near the critical point $X_c$ (where $X$ is one of the parameters \(\mathcal{J}^{(\nu)}\), \(\beta_\nu\), the other ones being fixed), the order parameter scales as  
$m \sim (X_c - X)^{\beta}$, with a critical exponent $\beta = 1/2$ (not to be confused with the inverse bath temperatures).
}

\APV{
The evaluation of $Z_{12}$ from Eq.~(\ref{eq:pstexp}) yields
\begin{equation}
Z_{12} = 
\sum_{\{s\}}
\exp\!\left\{
\frac{1}{4N}\left(\beta_1 \mathcal{J}^{(1)}+\beta_2 \mathcal{J}^{(2)}\right)
\left(\sum_{i=1}^{N}s_i\right)^2
+
\frac{1}{2}\left(\beta_1 h^{(1)}+\beta_2 h^{(2)}\right)\sum_{i=1}^{N}s_i
\right\}.
\end{equation}
Using the Gaussian identity  
$\int_{-\infty}^{\infty}\!\exp(-x^2 + 2ax)\,dx = \sqrt{\pi}\,e^{a^2}$,  
the expression above can be rewritten as
\[
Z_{12} =
\sum_{\{s\}}
\int_{-\infty}^{\infty}\!dx\,
\exp\!\left\{
-x^2 +
\left[
x \sqrt{\frac{\beta_1 \mathcal{J}^{(1)}+\beta_2 \mathcal{J}^{(2)}}{N}} +
\frac{1}{2}\left(\beta_1 h^{(1)}+\beta_2 h^{(2)}\right)
\right]\sum_{i=1}^{N}s_i
\right\}.
\]
This can be conveniently rewritten as
\begin{equation}
Z_{12} =
\sqrt{\frac{N\left(\beta_1 \mathcal{J}^{(1)}+\beta_2 \mathcal{J}^{(2)}\right)}{4\pi}}
\int_{-\infty}^{\infty}\!
dm\,
e^{-N\left(\beta_1 \mathcal{J}^{(1)}+\beta_2 \mathcal{J}^{(2)}\right)\,g_{12}(m)},
\end{equation}
where $m = \sum_{i=1}^{N}s_i/N$ and
\[
g_{12}(m)=
\frac{m^2}{2}
-
\frac{1}{\beta_1 \mathcal{J}^{(1)}+\beta_2 \mathcal{J}^{(2)}}
\ln\left\{
2\cosh\left[
\frac{1}{2}\left(\beta_1 \mathcal{J}^{(1)}+\beta_2 \mathcal{J}^{(2)}\right)m +
\frac{1}{2}\left(\beta_1 h^{(1)}+\beta_2 h^{(2)}\right)
\right]
\right\}.
\]
In the thermodynamic limit $N \rightarrow \infty$, the steady-state magnetization is associated with the maximum of $Z_{12}$, obtained by minimizing $g_{12}(m)$ with respect to $m$, from which Eq.~(\ref{mage}) is again recovered.
}

\APV{
As for the entropy production, it is given by  
$\langle \dot{\sigma} \rangle = -\beta_1 \langle \dot{Q}_1 \rangle - \beta_2 \langle \dot{Q}_2 \rangle$,  
where 
\begin{equation}
\langle \dot{Q}_\nu \rangle = \Delta E^{(\nu)}_{+-} \left(\omega_{+-}^{(\nu)}p_- -\omega_{-+}^{(\nu)}p_+  \right) + \Delta E^{(\nu)}_{-+} \left(\omega_{-+}^{(\nu)}p_+ -\omega_{+-}^{(\nu)}p_-  \right), 
\end{equation}
leading, after some manipulations, to Eq.~\eqref{eq:entprod} in the main text.
}

\section{
\APV{
The fast-alternation limit \(d\rightarrow\infty\) and finite-\(d\) corrections to the location of the transition point.
}
}\label{d}

\APV{
We employ a "two-box" description to model the case of a finite rate of alternation between the thermal baths with which the system is in contact \cite{danielPhysRevResearch.2.043257,liang2021dissipation,busiello2021dissipation}. This approach comprises two kinds of dynamics: intra-box evolution and switching between baths. In the former, the system is in contact with a single thermal bath $\nu$ at a time, and a configuration $s$ changes to $s'$ via a single spin flip, occurring with a rate $\omega^{(\nu)}_{s' s}$. The latter dynamic accounts for the switching between baths, $\nu\rightarrow \nu'$, with a rate $\Omega_{\nu' \nu}$ (which we will assume to be constant and equal to $d$), while the configuration $s$ of the system is held fixed.
}

\APV{
The time evolution for the probability of finding the system in state $s$ while in contact with bath $\nu$, ${p}^{(\nu)}_s(t)$, is given by
\begin{equation}\label{seq}
\dot{p}^{(\nu)}_s(t)=\sum_{s'\neq s}J^{(\nu)}_{s s'}(t)+\sum_{\nu'\neq \nu}\mathcal{K}^{(s)}_{\nu \nu'}(t),
\end{equation}
where the first term accounts for the intra-box dynamics, with the probability current $J^{(\nu)}_{s s'}(t)$ given by
\begin{equation}
J^{(\nu)}_{s s'}(t)= \omega^{(\nu)}_{s s'} p^{(\nu)}_{s'}(t) - \omega^{(\nu)}_{s' s} p^{(\nu)}_s(t),
\label{mee2}
\end{equation}
and the second term describes the switching between baths for a fixed state $s$, with
\begin{equation}
\mathcal{K}^{(s)}_{\nu \nu'}(t)=\Omega_{\nu \nu'} p^{(\nu')}_{s}(t)-\Omega_{\nu'\nu} p^{(\nu)}_s(t).
\end{equation}
}

\APV{
From the results in Ref.~\cite{danielPhysRevResearch.2.043257}, we can see that in the fast alternating limit \(d\rightarrow\infty\) (of ``simultaneous'' contact with the baths) the description can be simplified by using a coarse-grained probability \(p_s(t)=p^{(1)}_s(t)+p^{(2)}_s(t)\). Along with a proper rescaling of the remaining rates, this leads to Eq.~\eqref{eq:pAB}, with an entropy production given by Eq.~\eqref{ep} of the main text. 
}

\begin{figure}[h!]
\centering
\includegraphics[width=.6\textwidth]{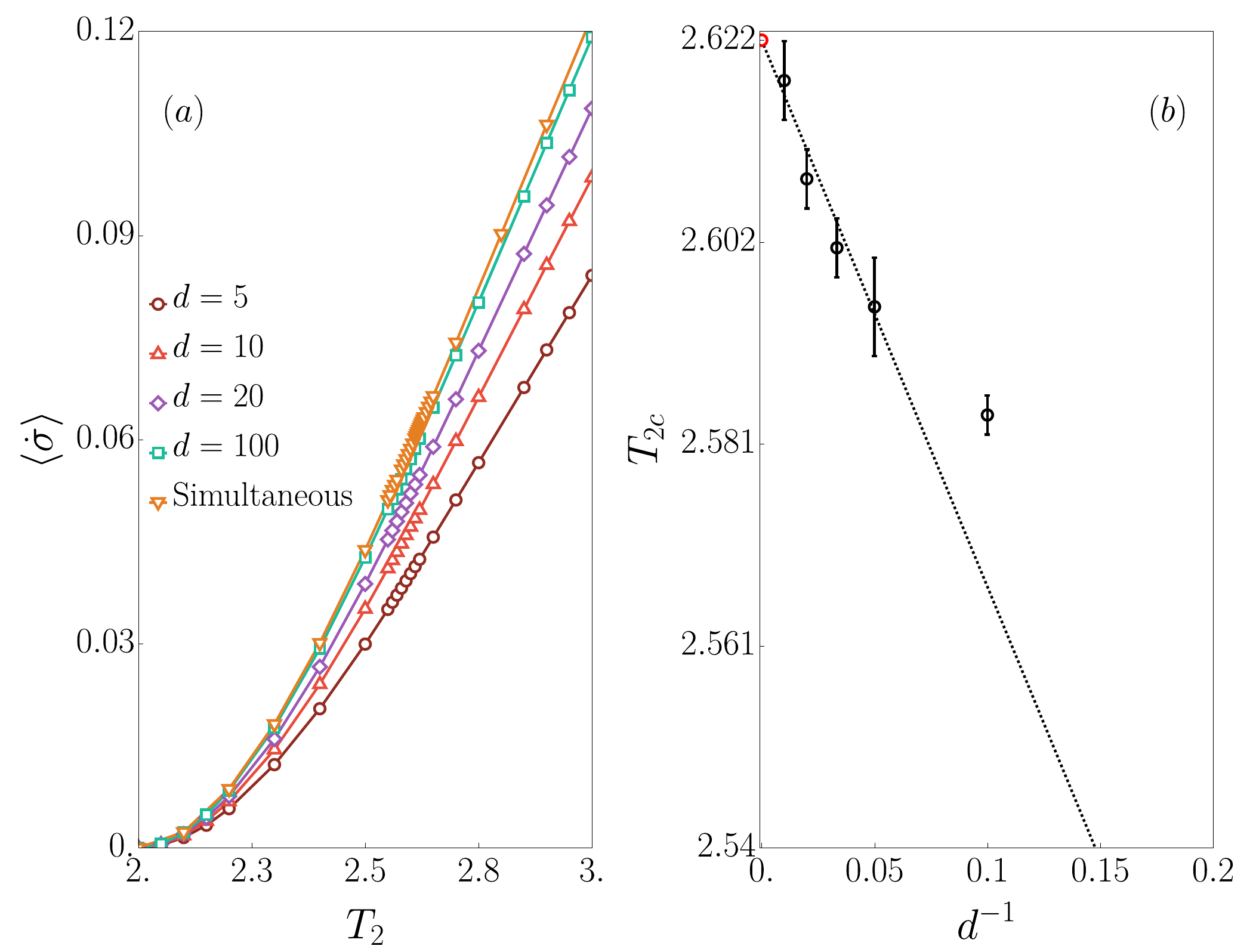}
\caption{
\APV{
For the same parameters as in Fig.~\ref{fig1} (main text), we plot (left) the entropy production $\langle {\dot \sigma}\rangle$ for different alternation rates $d$ (with $L=30$) and (right) the critical point $T_{2c}$ versus $d^{-1}$. The red symbol denotes the exact value for $d\rightarrow \infty$, which is equivalent to the simultaneous contact model. The straight line highlights the linear dependence on $d^{-1}$ for large $d$.}
}
\label{supp2}
\end{figure}

\APV{
In the mean-field limit, it can be shown analytically that the location of a transition point for finite \(d\) converges to that of the \(d\rightarrow\infty\) limit and that the corrections are at most of order \(1/d\) \cite{mamede2025collective}. This same behavior is observed in simulations, as illustrated for the square-lattice Ising model in Fig.~\ref{supp2}, which shows, using the same parameters as in Fig.~\ref{fig1} of the main text, the dependence of the entropy production $\mom{\dot{\sigma}}$ on  $T_2$ and the value of the critical temperature $T_{2c}$ as a function of the alternation rate $d$. The values of $T_{2c}$ were estimated from the crossing of Binder-cumulant curves for various system sizes. As $d$ increases, the thermodynamic quantities approach the values of the simultaneous contact case, and the critical temperature converges to the value $2/\left[2\ln\left(1+\sqrt{2}\right)-1\right]$ obtained from Eq.~\eqref{eqe}. For large $d$, the deviations again scale as $1/d$. We checked that this is also the typical behavior for the other models discussed in the main text.
}

\section{Main expressions for the all-to-all Potts model}\label{apd}

\APV{
Here we derive the main relations for the all-to-all Potts model. As in the case of the Ising model, it is more convenient to work with the fractions \(p_i=N_i/N\) of spins in the state \(i\in\{0,1,\dots,q-1\}\). The fraction \(p_0\) evolves in time according to the master equation 
\begin{equation}
\dot{p}_0(t)=\sum_{j=1}^{q-1}\left(\omega_{0j}^{(1)}+\omega_{0j}^{(2)}\right)p_j(t)
-\sum_{j=1}^{q-1}\left(\omega_{j0}^{(1)}+\omega_{j0}^{(2)}\right)p_0(t),
\end{equation}
in which \(\omega^{(\nu)}_{ij}=e^{-\frac{1}{2}\beta_\nu\Delta E^{(\nu)}_{ij}}\), with \(\Delta E^{(\nu)}_{ij}\) given by Eq.~\eqref{eq:DeltaEPotts}.
}

\APV{
In the NESS, the order parameter \(m\) is related to the set \(\{p_i\}\) through \(p_0=[1+(q-1)m]/q\) and \(p_j=(1-m)/q\) \((j=1,2,\dots,q-1)\). This leads to
\begin{equation}
0=
\left(e^{-\frac{1}{2}\beta_1\mathcal{J}^{(1)}m}+e^{-\frac{1}{2}\beta_2\mathcal{J}^{(2)}m}\right)(1-m)
-\left(e^{\frac{1}{2}\beta_1\mathcal{J}^{(1)}m}+e^{\frac{1}{2}\beta_2\mathcal{J}^{(2)}m}\right)[1+(q-1)m],
\label{order_Expansion_Genq}
\end{equation}
from which we obtain
\begin{equation}
m=\frac{2}{2-q+q\coth\left[\tfrac{1}{4}\left(\beta _1  \mathcal{J}^{(1)} +\beta _2  \mathcal{J}^{(2)}\right)m\right]}.
\label{PottsMinimum_MS}
\end{equation}
}

\APV{
To locate the coexistence lines for \(q>3\), we proceed as for the nonequilibrium Ising model in the limit \(N\to\infty\). The corresponding functional \( g_{12}(m)\) is given by
\begin{eqnarray}
g_{12}(m)&=&-\frac{1}{2}\left(\beta_1\mathcal{J}^{(1)}+\beta_2\mathcal{J}^{(2)}\right)\frac{1+(q-1)m^2}{2q}
+\left[\frac{1+(q-1)m}{q}\right]\ln\left[\frac{1+(q-1)m}{q}\right]
+\frac{(q-1)(1-m)}{q}\ln\left(\frac{1-m}{q}\right).
\end{eqnarray}
Expanding \(g_{12}(m)\) in powers of \(m\), we obtain
\[
g_{12}(m)= g_{12}(0)+\phi_2m^2-\phi_3m^3+\phi_4m^4+\cdots,
\]
the leading coefficients being given by
\begin{eqnarray}
g_{12}(0) &=& -\frac{\beta_1\mathcal{J}^{(1)}+\beta_2\mathcal{J}^{(2)}+4q\ln q}{4q},\nonumber\\
\phi_2 &=& \frac{q-1}{2q}\left[q-\frac{1}{2}\left(\beta_1\mathcal{J}^{(1)}+\beta_2\mathcal{J}^{(2)}\right)\right],\nonumber\\
\phi_3 &=& \frac{1}{6}(q-2)(q-1),\nonumber\\
\phi_4 &=& \frac{1}{12}(q-1)[(q-3)q+3].    
\end{eqnarray}
}

\APV{
For \(q=2\), \(\phi_3=0\) and the condition \(\phi_2=0\) defines a critical line \(\beta_1\mathcal{J}^{(1)} + \beta_2\mathcal{J}^{(2)} = 4\), equivalent to the result for the Ising model apart from a rescaling of the coupling constants by a factor of 2. For \(q\ge 3\), since \(\phi_3>0\), the transition is discontinuous. At phase coexistence, the conditions \(g_{12}(m_0)=g_{12}(0)\) and \(g_{12}^\prime(m_0)=0\) yield
\begin{equation}
\frac{1}{2}\left(\beta_1\mathcal{J}^{(1)}+\beta_2\mathcal{J}^{(2)}\right)=\frac{2(q-1)}{q-2}\ln(q-1),
\label{qGT2BetaV}
\end{equation}
as reported in the main text.
}

\APV{
From the order-parameter expression and the definition of the heat flux
\(\langle\dot{Q}_\nu\rangle\) [analogous to Eq.~\eqref{eq:Qnu}], we obtain
\begin{equation}
\langle\dot{Q}_\nu\rangle
=m\mathcal{J}^{(\nu)}\left\{e^{-\frac{1}{2}\beta_\nu\mathcal{J}^{(\nu)}m}(1-m) -e^{\frac{1}{2}\beta_\nu\mathcal{J}^{(\nu)}m}\left[1+(q-1)m\right]\right\},
\end{equation}
which can be rewritten as
\begin{equation}
\langle\dot{Q}_\nu\rangle =\frac{(q-1)m\mathcal{J}^{(\nu)}
\sinh\left[\tfrac{1}{4}(\beta _2 \mathcal{J}^{(2)}-\beta _1 \mathcal{J}^{(1)})m\right]}
{q e^{\frac{1}{4}\left(\beta _1 \mathcal{J}^{(1)}+\beta _2 \mathcal{J}^{(2)}\right)m}
-2\sinh\left[\frac{1}{4}\left(\beta_1 \mathcal{J}^{(1)}+\beta _2 \mathcal{J}^{(2)}\right)m\right]}.
\end{equation}
Finally, the entropy production \(\langle\dot{\sigma}\rangle=-\beta_1\langle\dot{Q}_1\rangle-\beta_2\langle\dot{Q}_2\rangle\) can be rewritten as
\begin{equation}
\langle\dot{\sigma}\rangle
=\frac{(q-1)\left(\beta _2 \mathcal{J}^{(2)}-\beta _1 \mathcal{J}^{(1)}\right)m
\sinh\left[\frac{1}{4}\left(\beta_2 \mathcal{J}^{(2)}-\beta _1\mathcal{J}^{(1)}\right)m\right]}
{q e^{\frac{1}{4}\left(\beta _1 \mathcal{J}^{(1)}+\beta _2 \mathcal{J}^{(2)}\right)m}
-2\sinh\left[\frac{1}{4}\left(\beta_1 \mathcal{J}^{(1)}+\beta _2 \mathcal{J}^{(2)}\right)m\right]}.
\end{equation}
%COMENTAR SOBRE COMPORTAMENTO DE ESCALA E SOBRE O FATO DE QUE A P.E. É NULA NA FASE DESORDENADA.
}

\section{
\APV{
Nonequilibrium Blume-Capel model and the steady-state solution for all-to-all interactions
}
}\label{apc}

\APV{
In the (mean-field) case of all-to-all interactions, the energies of the Blume-Capel model are given by Eq.~\eqref{eq:BCmf}. Once again, in the thermodynamic limit it is more convenient to the describe the dynamics in terms of the fractions \(p_0=N_0/N\) and \(p_\pm=N_\pm/N\) of spins in the individual states \(s_i=0\) and \(s_i=\pm1\). Taking into account that \(p_0=1-p_+-p_-\), the master equations for \(p_\pm\) are
\begin{eqnarray}
    \dot{p}_+ &=& f_+(p_+,p_-) = \left(\omega_{+0}^{(1)}+\omega_{+0}^{(2)}\right)(1-p_+-p_-) + \left(\omega_{+-}^{(1)}+\omega_{+-}^{(2)}\right)p_- - \left(\omega_{0+}^{(1)}+\omega_{0+}^{(2)}+\omega_{-+}^{(1)}+\omega_{-+}^{(2)}\right)p_+ ,\nonumber\\
    \dot{p}_- &=& f_-(p_+,p_-) = \left(\omega_{-+}^{(1)}+\omega_{-+}^{(2)}\right)p_+ + \left(\omega_{-0}^{(1)}+\omega_{-0}^{(2)}\right)(1-p_+-p_-) - \left(\omega_{0-}^{(1)}+\omega_{0-}^{(2)}+\omega_{+-}^{(1)}+\omega_{+-}^{(2)}\right)p_-,
    \label{eq:BCmemf}
\end{eqnarray}
in which \(\omega^{(\nu)}_{ij}=e^{-\frac{1}{2}\beta_\nu\Delta E^{(\nu)}_{ij}}\), with \(\Delta E^{(\nu)}_{ij}\), for \(i,j\in\{+,0,-\}\), given by Eqs.~\eqref{eq:BCmfDE}. Notice that \(\Delta E^{(\nu)}_{ij}\) depends on \(p_+\) and \(p_-\). Here we restrict ourselves to the zero-field case \(h^{(1)}=h^{(2)}=0\).
}

\APV{
The NESS, for which \(\dot{p}_+=\dot{p}_-=0\), always has a paramagnetic (disordered) solution \(p_+=p_-\equiv\frac{1}{2}\rho_\mathrm{para}\), in which \(\rho_\mathrm{para}\) is given by 
\begin{equation}
    \rho_\mathrm{para} = \frac{1}{1+\frac{1}{2} e^{\frac{1}{2}\left(\beta_1\Delta^{(1)}+\beta_2\Delta^{(2)}\right)}}.
\end{equation}
At sufficiently low temperatures, there must also exist an ordered solution. From any NESS solution for \(p_\pm\), we can obtain the magnetization \(m=p_+-p_-\) and the quadrupole moment \(\rho=p_++p_-\). In the disordered phase, we therefore have \(m=0\) and \(\rho=\rho_\mathrm{para}\). 
}

\APV{
Close to criticality, the NESS conditions can be used to express \(\rho\) as a series around \(m=0\) and to obtain an equation for \(m\) in terms of a series expansion of the form 
\begin{equation}
0=a_1 m + a_3 m^3 + a_5 m^5 + \mathcal{O}(m^7), 
\label{eq:mseriesBC}
\end{equation}
with
\begin{equation}
a_1 \propto \frac{1}{2}\left(\beta_1\mathcal{J}^{(1)}+\beta_{2}\mathcal{J}^{(2)}\right)-\left[1+\frac{1}{2}e^{\frac{1}{2}\left(\beta_1 \Delta^{(1)}+\beta_2 \Delta^{(2)}\right)}\right],
\end{equation}
with a coefficent \(a_3\) which may change sign and a coefficient \(a_5\) which is never zero for the points of interest here.
}

\APV{
The criticality condition corresponds to \(a_1=0\), with \(a_3\neq0\), leading to
\begin{equation}
\frac{1}{2}\left(\beta_1\mathcal{J}^{(1)}+\beta_{2}\mathcal{J}^{(2)}\right) = 1+\frac{1}{2}e^{\frac{1}{2}\left(\beta_1 \Delta^{(1)}+\beta_2 \Delta^{(2)}\right)},
\end{equation}
which reduces to the equilibrium expression, \( \beta \mathcal{J} = 1+ \frac{1}{2} e^{\beta\Delta}\) \cite{Capel}, when the two baths have the same parameters, in agreement with Eq.~\eqref{eqe}. 
}

\APV{
When \(a_1=0\), the coefficient \(a_3\) can be expressed as 
\begin{equation}
a_3 \propto 4-e^{\frac{1}{2}\left(\beta_1 \Delta^{(1)}+\beta_2 \Delta^{(2)}\right)}.
\end{equation}
A tricritical point corresponds to \(a_1=a_3=0\), leading to 
\begin{eqnarray}
    \frac{1}{2}\left(\beta_1\mathcal{J}^{(1)}+\beta_{2}\mathcal{J}^{(2)}\right) &=& 3, \\
    \frac{1}{2}\left(\beta_1 \Delta^{(1)}+\beta_2 \Delta^{(2)}\right) &=& \ln 4,
\end{eqnarray}
again in agreement with Eq.~\eqref{eqe}.
}

\APV{
Close to a critical point $X_c$ (where $X$ is one of the parameters \(\mathcal{J}^{(\nu)}\), \(\Delta^{(\nu)}\), \(\beta_\nu\), the other ones being fixed), we have \(a_1\propto|X_c - X|\) and \(a_3\neq0\), so that the order parameter scales as $m \sim |X_c - X|^{\beta_c}$, with a critical exponent $\beta_c = 1/2$ (not to be confused with the inverse bath temperatures). Analogously, close to a tricritical point $X_t$, with \(a_1\propto|X_t - X|\) and \(a_3=0\), we observe a scaling $m \sim |X_t - X|^{\beta_t}$, with a tricritical exponent $\beta_t = 1/4$. 
}

\APV{
The tricriticality condition can also be derived from Eq.~(\ref{eq:psinfH}). Analogously to the case of the Ising model, $Z_{12}$ can be written as
\begin{equation}
Z_{12}=\sum_{\{s\}}\exp\left\{\frac{1}{4N}\left(\beta_1\mathcal{J}^{(1)}+\beta_2\mathcal{J}^{(2)}\right)\left(\sum_{i=1}^Ns_i\right)^2 +\frac{1}{2}\left(\beta_1h^{(1)}+\beta^{(2)}h^{(2)}\right)\sum_{i=1}^Ns_i 
+\frac{1}{2}\left(\beta_1\Delta^{(1)}+\beta^{(2)}\Delta^{(2)}\right)\sum_{i=1}^Ns^2_i\right\},
\end{equation}
which is equivalent to
\begin{equation}
Z_{12}=\sqrt{\frac{N(\beta_1\mathcal{J}^{(1)}+\beta_2\mathcal{J}^{(2)})}{4\pi}} \int dm~d\rho~e^{-N g_{12}(\rho,m)}, 
\end{equation}
where
$g_{12}(m,\rho)$ is given by
%{
%As established in the main text, the steady-state probability distribution takes the form
%\begin{equation}
%    p^{\rm st}_{\bm{s}'} =\frac{e^{-\beta\overline{E}(\bm{s}')}}{Z},
%\end{equation}
%where $\beta\overline{E} \equiv [\beta_1 E^{(1)}(\bm{s}') + \beta_2 E^{(2)}(\bm{s}')]/2$ defines an effective energy. The free energy per site, $\mathbb{F} = %\beta\overline{\mathbb{E}} - \mathbb{S}$, follows from the thermodynamic limits
%\begin{equation}
%    \beta\overline{\mathbb{E}} = \lim_{N\to\infty} \beta\overline{E}/N, \quad
%    \mathbb{S} = \lim_{N\to\infty} S/N,
%\end{equation}
%representing the energy and entropy per site, respectively. For the BEG model, these %quantities in the $N\to\infty$ limit are given by
%\begin{align}
%    {\tilde g}_{12}&=\frac{1}{2} \left(\beta _1 h^{(1)}+\beta _2 %h^{(2)}\right) \left(p_{+}-p_{-}\right)-\frac{1}{4} (\beta _1 %\mathcal{J}^{(1)}+\beta _2 \mathcal{J}^{(2)}) \left(p_{+}-p_{-}\right)%{}^2+\frac{1}{2} \left(\beta _1 \Delta _1+\beta _2 \Delta _2\right) %\left(p_{-}+p_{+}\right)\nonumber\\&+p_{+}\ln p_{+}+ p_{-}\ln  p_{-}+(1- %p_{+}- p_{-})\ln(1- p_{+}- p_{-}).
%\end{align}
%}
%By expressing ${\tilde g}_{12}$ in terms 
%of $m = p_{+} - p_{-}$ and
%$\rho = p_{+} + p_{-}$, one has that
\begin{equation}
    g_{12}(m,\rho) = -\frac{1}{4}( \beta_1\mathcal{J}^{(1)} + \beta_2\mathcal{J}^{(2)}) m^2 + \frac{1}{2}(\beta_1\Delta^{(1)} + \beta_2\Delta^{(2)})\rho + \frac{1}{2}(\beta_1h^{(1)} + \beta_2h^{(2)})m + S(m,\rho),
\end{equation}
with
\begin{equation}
    S(m,\rho) = \frac{\rho - m}{2} \ln\left(\frac{\rho - m}{2}\right) 
    + \frac{\rho + m}{2} \ln\left(\frac{\rho + m}{2}\right) 
    + (1 - \rho) \ln(1 - \rho),
\end{equation}
By minimizing $g_{12}(m,\rho)$ with respect to $m$ and $\rho$ we arrive at the equations of state
\begin{equation}
    \mathcal{K} m = \mathcal{B} + \ln\left(\frac{\rho + m}{\rho - m}\right), \quad
    \mathcal{D} = \ln\left(\frac{4(1 - \rho)^2}{q^2 - m^2}\right).
    \label{sec}
\end{equation}
where
\begin{equation}
    \mathcal{K} \equiv \beta_1\mathcal{J}^{(1)} + \beta_2\mathcal{J}^{(2)}, \quad
    \mathcal{D} \equiv \beta_1\Delta^{(1)} + \beta_2\Delta^{(2)}, \quad
    \mathcal{B} \equiv \beta_1h^{(1)} + \beta_2h^{(2)},
\end{equation}
}

\APV{
The second equation in (\ref{sec})  can be rewritten as
\begin{equation}
    \rho(m) = \frac{e^{\mathcal{D}/2} \sqrt{(e^{\mathcal{D}} - 4)m^2 + 4} - 4}{e^{\mathcal{D}} - 4}.
\end{equation}
Inserting this last result into the expression for $g_{12}(m,\rho)$, specializing for the zero-field case $ \beta_1 h^{(1)} + \beta_2 h^{(2)} = 0$, and expanding for small $m$ we arrive at 
\begin{equation}
    g(m)\equiv g_{12}\left(m,\rho(m)\right) \approx g(0) + \phi_2 m^2 + \phi_4 m^4,
\end{equation}
where
\begin{align}
    g(0) &= \frac{1}{2} \left[\mathcal{D} - 2 \ln\left(e^{\mathcal{D}/2} + 2\right)\right], \\
%    \phi_1 &= \frac{\mathcal{B}}{2}, \\
    \phi_2 &= \frac{1}{4}\left(e^{\mathcal{D}/2} + 2 - \mathcal{K}\right), \\
    \phi_4 &= -\frac{1}{192} \left(e^{\mathcal{D}/2} - 4\right) \left(e^{\mathcal{D}/2} + 2\right)^2.
\end{align}
The order-disorder transition occurs for $\phi_2 = 0$, leading to the criticality condition
\begin{equation}
    \frac{1}{2}\left(\beta_1 \mathcal{J}^{(1)} + \beta_2 \mathcal{J}^{(2)}\right) = 1 + \frac{1}{2} e^{\frac{1}{2}(\beta_1 \Delta^{(1)} + \beta_2 \Delta^{(2)})}.
\end{equation}
Moreover, a tricritical point emerges when $\phi_2=\phi_4=0$, leading to the conditions
\begin{equation}
    \frac{1}{2}\left(\beta_1 \Delta^{(1)} + \beta_2 \Delta^{(2)}\right)= \ln 4 \qquad {\text and}\qquad \frac{1}{2}\left(\beta_1\mathcal{J}^{(1)} + \beta_2 \mathcal{J}^{(2)}\right) = 3,
\end{equation}
in agreement with results shown in the main text from the order-parameter expansion via the master equation.
}

\APV{
In the ordered phase, close to a critical point, the entropy production \(\langle\dot\sigma\rangle\) can be expressed as a series in the magnetization, yielding
\begin{equation}
\langle\dot\sigma\rangle = b_0 + b_2 m^2+b_4 m^4+\mathcal{O}(m^6),
\end{equation}
with (generically) nonzero coefficients \(b_2\) and \(b_4\), while \(b_0\) can be written as its value \(\langle\dot\sigma\rangle_c\) at the critical point plus a term proportional to the distance to criticality \(g\propto(X_i-X)\). (Full expressions for the coefficients are lengthy and not very informative.) Therefore, close to a generic criticality condition, i.e. one at which \(\beta_1\mathcal{J}^{(1)}\neq \beta_2\mathcal{J}^{(2)}\) and \(\beta_1\Delta^{(1)}\neq \beta_2\Delta^{(2)}\), the scaling of \(\langle\dot\sigma\rangle\) in the ordered phase around a critical (\(i=c\)) or tricritical (\(i=t)\) point follows
\begin{equation}
\langle\dot\sigma\rangle-\langle\dot\sigma\rangle_c\sim|g|^{1-\zeta_i},
\end{equation}
with \(\zeta_i=1-2\beta_i\), leading to \( \zeta_t = 1/2 \) at a tricritical point and \( \zeta_c = 0 \) at a critical point, in the last case with a discontinuous derivative \(d\mom{\dot{\sigma}}/dg\). The value \(\zeta_c=0\) is the same as in the case of the Ising model under field combination (I), and it is also observed in the systems investigated in Ref.~\cite{noa2019}. On the other hand, under ``independent'' criticality conditions, at which \(\beta_1\mathcal{J}^{(1)}= \beta_2\mathcal{J}^{(2)}\) and \(\beta_1\Delta^{(1)}= \beta_2\Delta^{(2)}\), a careful analysis points to \(\zeta_c = 4\) but still \(\zeta_t = 1/2\).  
}

\APV{
In the paramagnetic (disordered) phase, \(\langle\dot\sigma\rangle\) is given by
\begin{equation}
   \langle {\dot \sigma} \rangle_\mathrm{para} = 2\left(\beta_1 \Delta^{(1)}-\beta_2 \Delta^{(2)}\right) \frac{e^{\frac{1}{2}\beta_1 \Delta^{(1)}}-e^{\frac{1}{2}\beta_2 \Delta^{(2)}}}{1+\frac{1}{2}e^{\frac{1}{2}\left(\beta_1 \Delta^{(1)}+\beta_2 \Delta^{(2)}\right)}},
\end{equation}
and, contrary to what is verified for the Potts model, \(\langle {\dot \sigma} \rangle_\mathrm{para}\) depends on the distance to the transition point.
}

\end{document}